\documentclass[aps,prd,onecolumn,twoside,notitlepage,preprintnumbers,showpacs,
showkeys,superscriptaddress,byrevtex,floatfix,nofootinbib,amsmath,amssymb,longbibliography]{revtex4-1}

\usepackage{graphicx}
\usepackage{dcolumn}
\usepackage{bm}
\usepackage{tensor}
\usepackage{cancel}
\usepackage{array,amscd,amsmath,amssymb,amsthm,graphicx,mathtools,placeins,slashed,verbatim,tensor,dsfont,enumitem}
\usepackage[bookmarksnumbered,bookmarksopen=true,bookmarksopenlevel=1,pdfstartview=FitH,hidelinks]{hyperref}
\usepackage[all]{hypcap}
\usepackage{rotating}
\usepackage[caption=false]{subfig}
\usepackage{eepic}
\usepackage{braket}
\usepackage{mathrsfs}

\newcommand{\R}{\mathds{R}}
\newcommand{\C}{\mathds{C}}
\newcommand{\Q}{\mathds{H}}
\newcommand{\Oct}{\mathds{O}}
\newcommand{\Al}{\mathds{A}}
\newcommand{\id}{\mathds{1}}

\newcommand{\charge}{\mathcal{Q}}

\newcommand{\rep}[1]{\mathbf{#1}}
\newcommand{\brep}[1]{\mathbf{\overline{#1}}}

\newcolumntype{L}[1]{>{\raggedright\let\newline\\\arraybackslash\hspace{0pt}}m{#1}}
\newcolumntype{C}[1]{>{\centering\let\newline\\\arraybackslash\hspace{0pt}}m{#1}}
\newcolumntype{R}[1]{>{\raggedLeft\let\newline\\\arraybackslash\hspace{0pt}}m{#1}}

\newcommand{\be}{\begin{equation}}
\newcommand{\ee}{\end{equation}}
\newcommand{\susy}{\mathcal{N}}

\newcommand{\N}{\ensuremath{\mathcal{N}}}

\newcommand{\aref}[1]{\hyperref[#1]{appendix~\ref{#1}}}
\begin{document}

\preprint{DIAS-STP-17-06, IMPERIAL-TP-2017-MJD-01, DFPD/2017/TH-08, NORDITA 2017-067}

\title{Are all supergravity theories Yang-Mills squared?}

\author{A. Anastasiou}
\email[]{alexandros.anastasiou@su.se}
\affiliation{Nordita, KTH Royal Institute of Technology and Stockholm University, Roslagstullsbacken 23, 10691 Stockholm, Sweden}
\author{L. Borsten}
\email[]{leron@stp.dias.ie}
\affiliation{School of Theoretical Physics, Dublin Institute for Advanced Studies,
10 Burlington Road, Dublin 4, Ireland}
\author{M. J. Duff}
\email[]{m.duff@imperial.ac.uk}
\affiliation{Theoretical Physics, Blackett Laboratory, Imperial College London,
London SW7 2AZ, United Kingdom}
\affiliation{Mathematical Institute, University of Oxford, Andrew Wiles Building, Woodstock Road, Radcliffe Observatory Quarter,
Oxford, OX2 6GG, United Kingdom}
\author{A. Marrani}
\email[]{alessio.marrani@pd.infn.it}
\affiliation{Centro Studi e Ricerche ``Enrico Fermi'',
Via Panisperna 89A, I-00184, Roma, Italy}
\affiliation{Dipartimento di Fisica e Astronomia ``Galileo Galilei'', Universit\`a di Padova, and INFN, sezione di
Padova,
Via Marzolo 8, I-35131 Padova, Italy}
\author{S. Nagy}
\email[]{silvia.nagy1@nottingham.ac.uk}
\affiliation{Centre for Astronomy \& Particle Theory,
University Park,
Nottingham,
NG7 2RD,
United Kingdom}
\affiliation{ Center for Mathematical Analysis, Geometry and Dynamical Systems,
  Department of Mathematics, 
  Instituto Superior T\'ecnico, Universidade de Lisboa,
  Av. Rovisco Pais, 1049-001 Lisboa, Portugal}
\author{M. Zoccali}
\email[]{m.zoccali14@imperial.ac.uk}
\affiliation{Theoretical Physics, Blackett Laboratory, Imperial College London,
London SW7 2AZ, United Kingdom}
\date{\today}

\begin{abstract}
Using simple symmetry arguments we classify the ungauged $D=4$,     $\mathcal{N}=2$ supergravity theories, coupled to both vector and hyper multiplets through homogeneous scalar manifolds, that can be built as the product  of $\mathcal{N}=2$ and  $\mathcal{N}=0$ matter-coupled Yang-Mills gauge theories. This includes all such supergravities with two  isolated exceptions:  pure supergravity and the $T^3$ model. 
\end{abstract}


\maketitle

\makeatletter
\def\l@subsubsection#1#2{}
\makeatother

\tableofcontents

\newpage

\section{Introduction}


A   field theoretic incarnation  of  ``gravity $=$ gauge $\times$ gauge'' theory was developed in \cite{Anastasiou:2014qba, Nagy:2014jza, Anastasiou:2015vba, Borsten:2015pla,Anastasiou:2016csv, Cardoso:2016ngt, Cardoso:2016amd}. In particular,    one can form the product of fields belonging to two independent (super) Yang-Mills gauge theories, which we will refer to as the \emph{Left} and \emph{Right} factors. Importantly, the product maps the  content,  symmetries and field equations of the factors into those of a (super) gravity theory. We will refer to this construction here as \emph{squaring} Yang-Mills. In the present contribution  we use this framework to classify the ungauged $D=4$,     $\mathcal{N}=2$ supergravity theories, coupled to \emph{both vector and hyper multiplets} through homogeneous scalar manifolds, that can be built as the square of Yang-Mills.\\
\\
A  prior, related but distinct, realisation of the gravity $=$ gauge $\times$ gauge picture is given by the Bern-Carrasco-Johansson (BCJ) \emph{double-copy}  construction of scattering amplitudes. It has been conjectured \cite{Bern:2010yg,Bern:2010ue}, with  substantial evidence \cite{Bern:2009kd, Bern:2012cd, Bern:2013uka,  Bern:2013yya, Bern:2013qca,Bern:2014lha, Bern:2014sna}, that the scattering amplitudes of certain gravity theories are the double-copy, in a precise sense,  of amplitudes belonging to two independent Yang-Mills  theories. The paradigmatic example is given by $\mathcal{N}=8$ supergravity as the product of two $\mathcal{N}=4$ Yang-Mills theories, which due to the high degree of symmetry is  also the simplest possible case. These remarkable  amplitude relations rely crucially on the Bern-Carrasco-Johansson (BCJ) colour-kinematic duality \cite{Bern:2008qj}, which has been established at tree-level but remains conjectural for arbitrary loops, and have been used to demonstrate the existence  of unexpected cancellations throwing open the possibility that $\mathcal{N}=8$ supergravity may be perturbatively finite \cite{Bern:2012cd, Bern:2014sna}. There is now a growing list of double-copy constructible  theories in diverse dimensions \cite{Carrasco:2012ca, Damgaard:2012fb, Huang:2012wr, Bargheer:2012gv, Johansson:2014zca, Chiodaroli:2014xia, Chiodaroli:2015rdg, Chiodaroli:2015wal, Chiodaroli:2016jqw, Johansson:2017bfl},  conformal gravity being the latest addition \cite{Johansson:2017srf, Azevedo:2017lkz}. Moreover,  the BCJ amplitude prescription has  recently  been generalised to  include certain  curved background spacetimes \cite{Adamo:2017nia}.  At the same time, the paradigm has been extended beyond amplitudes in a variety of directions \cite{Monteiro:2011pc,BjerrumBohr:2012mg,Monteiro:2013rya,Fu:2016plh,  Anastasiou:2014qba, Nagy:2014jza, Anastasiou:2015vba, Borsten:2015pla,Monteiro:2014cda, Luna:2015paa, Luna:2016due, White:2016jzc, Goldberger:2016iau, Cardoso:2016ngt, Cardoso:2016amd, Goldberger:2017frp}. These remarkable and continually developing relations raise three natural questions: 
\begin{enumerate}[label=\roman*)]
\item Why does the correspondence work? Can we prove the BCJ colour-kinematic conjecture and pinpoint its origins?
\item How deep is the correspondence? That is, how far beyond amplitude relations can it be taken?
\item  How general is the correspondence? What gravitational theories admit a Yang-Mills  squared origin; are the factorisable theories special in some regard?
\end{enumerate}
Here we address a corner of (iii), by significantly extending the domain of ungauged $D=4$, $\mathcal{N}=2$ supergravity theories that are the square of Yang-Mills and hence may be in principle double-copy constructible.\\
\\
We should be clear about our definition of squaring: the gravitational theory   is \emph{defined} by the \emph{totality}  generated by the two gauge factors\footnote{Totality here includes everything generated by the product of adjoint multiplets with adjoint multiplets and fundamental multiplets with fundamental multiplets, but we exclude the product of adjoint multiplets with fundamental multiplets. This is reflected by the block diagonal spectator scalar in \eqref{spec}. Although a choice, since we seek generic statements and the adjoint-fundamental products are only consistent in very special circumstances (both in terms of  our ``square'' and the BCJ double-copy) it is actually imposed on us.}. In terms of the  double-copy this implies: (1) all gravity  scattering amplitudes can be factorised, in the BCJ  sense, into the product of amplitudes of the two gauge theories and (2) all double-copies of  the gauge theory amplitudes generate an amplitude belonging to the corresponding gravitational theory\footnote{As it stands this can of course only be established in general at tree-level, with supporting evidence  from case-by-case examples of loop-level amplitudes. Our present analysis is explicitly tree-level only.}. For example,  pure Einstein gravity is \emph{not} double-copy constructible in this sense.  Although all its amplitudes may be systematically double-copy constructed by consistently cancelling the would-be axion-dilaton sector with the product of ``ghost'' chiral fermion amplitudes \cite{Johansson:2014zca}, thus satisfying (1),  the spin-1 states arising in the ghost $\times$ ghost sector must be  explicitly (but consistently) excluded, thus failing (2).\\
\\
In the attempt to classify all supergravity theories with a Yang-Mills origin, the squaring and double-copy approaches are complementary in the following sense: starting with the double-copy, one finds the most general BCJ-friendly Yang-Mills candidate factors, then double-copies the amplitudes and lists the supergravity theories generated. Demanding BCJ duality  constrains  the couplings and symmetries of the gauge factors and one should be able to check that the resulting supergravities have the expected symmetries in the squaring sense. Alternatively, starting with squaring, one studies case-by-case whether or not each known supergravity theory  admits a factorisation using symmetry  principles, and only then checks for BCJ compatibility. These complementary pictures  have led to a  good understanding of a large  subset of gravity theories: for pure super Yang-Mills factors we have a complete classification of all supergravity theories generated for spacetime dimensions $3\leq D\leq 10$ \cite{Damgaard:2012fb, Borsten:2013bp, Anastasiou:2013hba, Anastasiou:2015vba}; using a factorisable orbifold construction and an $\mathcal{N}=0$ Yang-Mills factor, this was generalised to include a number of additional $\mathcal{N}=4,2,1$ matter-coupled supergravity theories \cite{Carrasco:2012ca}; in \cite{Johansson:2014zca} the colour-kinematic duality was generalised to include non-adjoint representations of the gauge group, allowing for fundamental matter-coupled Yang-Mills factors and a broader class of matter-coupled gravity theories;   this was subsequently used to double-copy construct all  ungauged $D=5, 4, \mathcal{N}=2$ supergravity theories coupled to  vector multiplets through a homogeneous scalar manifold, using half-hyper multiplets carrying a pseudo-real gauge group representation \cite{Chiodaroli:2015wal}. Building on such  principles, the symmetry arguments used in the present work were developed to construct all \emph{twin} supergravities in \cite{Anastasiou:2016csv}. In summary, so far the classification includes all  greater than half-maximal supergravities, all the half-maximal supergravities coupled to vector multiplets, some of the quarter maximal supergravities coupled to vector multiplets  and a small set of  simple theories outside these classes. Note,  in all  cases treated thus far the  scalars  parametrise  a homogeneous manifold, however, for $\N\leq 2$, supergravity theories with non-homogeneous scalar manifolds are also possible. The square or double-copy origin of such theories remains a compelling open question.\\
\\
Here, we adopt the squaring methodology to extend the domain of \cite{Damgaard:2012fb, Borsten:2013bp, Anastasiou:2013hba, Anastasiou:2015vba, Carrasco:2012ca, Johansson:2014zca, Chiodaroli:2015wal, Anastasiou:2016csv} to include $\N=2$ ungauged supergravity theories coupled to  both vector and hyper multiplets with homogeneous scalar manifolds using an $\N_L=2$ Yang-Mills theory coupled to a single half-hyper multiplet and a unique  class of $\N_R=0$ Yang-Mills gauge theories parametrised by six integers.
It is shown that, with the single exception of the $T^3$ model and pure $\N=2$ supergravity, the classification includes \emph{all}  such supergravity theories with symmetric scalar cosets.  For the non-symmetric  theories including hyper multiplets, we propose a candidate squaring procedure; there is a seemingly unique possibility involving a restriction to a  diagonal subgroup of the Left and Right global symmetries. Although  the origin of such a restriction remains unclear, our analysis suggests  that any $[\N_L=2] \times [\N_R=0]$ double-copy amplitude construction will reflect this requirement. Note, since the hyper multiplets  are insensitive to dimensional reduction and  it is only the scalar fields of the Right theory that contribute to this sector, our construction generalises trivially in all cases to $D=6,5,4,3$. \\
\\
\begin{table}[h!]
\centering
\begin{tabular}{C{1cm} C{2cm} C{1.5cm} C{2cm} L{10cm}}
\hline\hline
\\
~~$\charge$~~ & ~~$R$~~ & ~~$\text{Type}_{\susy}$~~ & ~~$f$~~ & ~~Content under $U(1)^{st}\times R$~~
\\ \\
\hline
\\
~~$32$~~ & ~~$SU(8)$~~ & ~~$\rep{G}_8$~~ & ~~$256$~~
& ~~$\rep{1_{-4}+8_{-3}+28_{-2}+56_{-1}+70_0+\overline{56}_1+\overline{28}_2+\overline{8}_3+1_{4}}$~~
\\
~~$28$~~ & ~~$U(7)$~~ & ~~$\rep{G}_7$~~ & ~~$256$~~
& ~~$\rep{1_{-4}^0+7_{-3}^1+1_{-3}^{-7}+21_{-2}^2+7_{-2}^{-6}+35_{-1}^3+21_{-1}^{-5}+\overline{35}_0^4+c.c.}$~~
\\
~~$24$~~ & ~~$U(6)$~~ & ~~$\rep{G}_6$~~ & ~~$128$~~
& ~~$\rep{1_{-4}^0+6_{-3}^1+15_{-2}^2+1_{-2}^{-6}+20_{-1}^{3}+6_{-1}^{-5}+\overline{15}_0^{4}+c.c.}$~~
\\
~~$20$~~ & ~~$U(5)$~~ & ~~$\rep{G}_5$~~ & ~~$64$~~
& ~~$\rep{1_{-4}^0+5_{-3}^1+10_{-2}^2+\overline{10}_{-1}^3+1_{-1}^{-5}+\overline{5}_0^{4}+c.c.}$~~
\\
~~$16$~~ & ~~$U(4)$~~ & ~~$\rep{G}_4$~~ & ~~$32$~~
& ~~$\rep{1_{-4}^0+4_{-3}^1+6_{-2}^2+\overline{4}_{-1}^3+1_0^{4}+c.c.}$~~
\\
~~$16$~~ & ~~$SU(4)$~~ & ~~$\rep{V}_4$~~ & ~~$16$~~
& ~~$\rep{1_{-2}+4_{-1}+6_{0}+\overline{4}_{1}+1_2}$~~
\\
~~$12$~~ & ~~$U(3)$~~ & ~~$\rep{G}_3$~~ & ~~$16$~~
& ~~$\rep{1_{-4}^0+3_{-3}^1+\overline{3}_{-2}^2+1_{-1}^3+c.c.}$~~
\\
~~$12$~~ & ~~$U(3)$~~ & ~~$\rep{V}_3$~~ & ~~$16$~~
& ~~$\rep{1_{-2}^0+3_{-1}^1+1_{-1}^{-3}+\overline{3}_{0}^2+c.c.}$~~
\\
~~$8$~~ & ~~$U(2)$~~ & ~~$\rep{G}_2$~~ & ~~$8$~~
& ~~$\rep{1_{-4}^0+2_{-3}^1+1_{-2}^2+c.c.}$~~
\\
~~$8$~~ & ~~$U(2)$~~ & ~~$\rep{V}_2$~~ & ~~$8$~~
& ~~$\rep{1_{-2}^0+2_{-1}^1+1_{0}^2+c.c.}$~~
\\
~~$8$~~ & ~~$U(2)$~~ & ~~$\rep{H}_2$~~ & ~~$8$~~
& ~~$\rep{1_{-1}^r+2_{0}^{r+1}+1_{1}^{r+2}+c.c.}$~~
\\
~~$8$~~ & ~~$U(2)$~~ & ~~$\rep{C}_2$~~ & ~~$4$~~
& ~~$\rep{1^{-1}_{-1}+2^{0}_{0}+1^{1}_{1}}$~~
\\
~~$4$~~ & ~~$U(1)$~~ & ~~$\rep{G}_1$~~ & ~~$4$~~
& ~~$(-4,0)+(-3,1)+c.c.$~~
\\
~~$4$~~ & ~~$U(1)$~~ & ~~$\rep{V}_1$~~ & ~~$4$~~
& ~~$(-2,0)+(-1,1)+c.c.$~~
\\
~~$4$~~ & ~~$U(1)$~~ & ~~$\rep{C}_1$~~ & ~~$4$~~
& ~~$(-1,r)+(0,r+1)+c.c.$~~
\\
~~$0$~~ & ~~-//-~~ & ~~$A$~~ & ~~$2$~~
& ~~$(-2)+c.c.$~~
\\
~~$0$~~ & ~~-//-~~ & ~~$\lambda$~~ & ~~$2$~~
& ~~$(-1)+c.c.$~~
\\
~~$0$~~ & ~~-//-~~ & ~~$\phi$~~ & ~~$2$~~
& ~~$(0)$~~
\\ \\
\hline\hline
\end{tabular}
\caption{On-shell helicity states of all $D=4$ supermultiplets. Here $\mathcal{Q}$ counts the number of supercharges, $R$ denotes the global $R$-symmetry group, $\text{Type}_\mathcal{N}$ the class of $\mathcal{N}$-extended supermultiplet and $f$ is  number of degrees of freedom. The $\mathcal{N}$-extended gravity, vector and spinor multiplets are denoted by $\rep{G}_\mathcal{N}, \rep{V}_\mathcal{N}$ and $\rep{C}_\mathcal{N}$, respectively. Note, $\rep{C}_{2}$ and $\rep{H}_{2}$ are used to distinguish half-hyper and full-hyper multiplets, respectively. Although $\rep{V}_3$ and   $\rep{V}_4$ are identical as isolated gauge multiplets, when coupled to supergravity they must be distinguished. Similarly, $\rep{G}_7$ and   $\rep{G}_8$ have identical content and as interacting theories are identical despite having \emph{a priori} distinct symmetries. Finally, we use $A, \lambda$ and $\phi$ to denote the smallest $\mathcal{N}=0$ vector, spinor and scalar multiplets, respectively. Sub/superscripts in the final column  refer to the $U(1)$ charges carried by the representations. The subscripts refer to the $U(1)^{st}$ helicities, which we  uniformly multiply by a factor of two for notational clarity. The superscripts refer to the internal $U(1)$ charges. When the symmetry has no semi-simple part (i.e.~for $\charge\leq 4$) we use  tuplets $(a, b, c,\ldots)$ to label the $U(1)$ charges, with the first slot reserved for  $U(1)_{st}$.}\label{smults}
\end{table}

\noindent The remaining sections are organised as follows. In \autoref{sugras} we summarise the class of supergravity theories considered here. In \autoref{squares} we consider the Yang-Mills  origin of these theories. We first outline the general principles in \autoref{genpric}. Then  we include only  vector multiplet couplings in \autoref{vectors}, which builds on the set of supergravities derived in     \cite{Chiodaroli:2015wal} by including a detailed analysis of the minimally coupled sequence and the $T^3$ model.  Finally, in \autoref{hypers} we include  the most general hyper multiplet couplings  by allowing for two independent fundamental scalars in the Right gauge theory factor. Before moving on, to fix our notation we summarise in \autoref{smults} the on-shell helicity states of all $D=4$ supermultiplets.

\section{$\mathcal{N}=2$ Supergravity}\label{sugras}

In this section we itemize the $D=4$,  $\mathcal{N}=2$  supergravity theories under consideration, specifically those with scalar fields parametrising a homogeneous manifold, highlighting their field content and symmetries as required for the Yang-Mills squared construction given in \autoref{squares}. We consider both vector and hyper multiplets; the total homogeneous scalar manifold $\mathcal{M}$ factors into a special K\"ahler (SK) manifold  ${G}/{H}$, parametrised by the scalars belonging to the vector multiplets, and a quaternionic (Q) manifold  $\mathcal{G/H}$, parametrised by the scalars belonging to the hyper multiplets,
\be\label{totcoset}
\mathcal{M}\cong \frac{G}{H}\times  \frac{\mathcal{G}}{\mathcal{H}}.
\ee
In \autoref{vector} and \autoref{hyper} we present the possible couplings to vector and hyper multiplets, respectively, under the assumption that the scalar manifold is homogeneous.

\subsection{Vector multiplets}\label{vector}

When coupling $\mathcal{N}=2$ supergravity to vector multiplets the scalar manifold must be projective SK \cite{deWit:1983xhu, deWit:1984wbb, Strominger:1990pd}. In the  non-symmetric case  the possible classes of  scalar manifolds are  indexed by three integers $(q, P, \dot{P})$ as described in \autoref{nonsym}. If the scalar manifold is symmetric there are three classes: $(i)$ the generic Jordan sequence indexed by a single integer, $(q, P, \dot{P})=(q, 0, 0)$, $(ii)$ the four magic supergravities \cite{Gunaydin:1983bi, Gunaydin:1983rk, Gunaydin:1984ak} for which  $(q, P, \dot{P})=(n, 1, 0)$, where $n=\dim \mathds{A}=1,2,4,8$, and  $(iii)$ the minimally coupled sequence indexed by a single integer, $(q, P, \dot{P})=(-2, P, 0)$. In addition to these classes, we have the isolated case of the $T^3$ model \cite{deWit:1995tf, Saraikin:2007jc}, which although underpinned by a Jordan algebra is not a part of the generic Jordan sequence \cite{Borsten:2011nq, Borsten:2011ai}. In the absence of hyper scalars $\mathcal{G}/\mathcal{H}$ reduces to a trivial $SU(2)/SU(2)$ factor, where the denominator corresponds to the global R-symmetry. 

\subsubsection{Non-symmetric}\label{nonsym}

In the non-symmetric homogeneous cases the scalar manifold is:
\be\label{SKnon}
\frac{G}{H}\times\frac{SU(2)}{SU(2)}=SO(1,1)\times\frac{SO(q+2,2)}{SO(q+2)\times U(1)}\times\frac{S_{q}(P,\dot{P})}{S_{q}(P,\dot{P})}\times\frac{SU(2)}{SU(2)}\ltimes\Big[(\rep{spin},\rep{def},\rep{1})^1\ltimes(\rep{1,1,1})^2\Big],
\ee
where $\rep{spin}$ indicates the spinor representation of $SO(q+2,2)$ and $\rep{def}$ the defining representation of $S_q(P,  \dot{P})$. Here, $(q, P,\dot{P})$ are integers, which fix the number of vector multiplets present, the symmetry groups and representations carried by the field content, as described in \autoref{SqPPreps}.\\
\begin{table}[h!]
\begin{tabular}{C{1cm} C{3cm} C{3cm} C{4cm} C{4cm} C{2cm}}
\hline\hline
\\
~~$q$~~ & ~~$SO(q+2)$~~ & ~~$S_q(P,\dot{P})$~~ & ~~$\rep{r}$~~ & ~~$\brep{r}$~~ & ~~$r(q,P,\dot{P})$~~
\\ \\
\hline \\
~~$-1$~~ & ~~-//-~~ & ~~$SO(P)$~~ & ~~$\rep{P}$~~ & ~~$\rep{P}$~~ & ~~$P$~~
\\
~~$0$~~ & ~~$U(1)$~~ & ~~$SO(P)\times SO(\dot{P})$~~ & ~~$(\rep{P}, \rep{1})^{-x}+(\rep{1}, \rep{\dot{P}})^{-y}$~~ & ~~$(\rep{P}, \rep{1})^{x}+(\rep{1}, \rep{\dot{P}})^{y}$~~ & ~~$P+\dot{P}$~~
\\
~~$1$~~ & ~~$SU(2)$~~ & ~~$SO(P)$~~ & ~~$\rep{(2,P)}$~~ & ~~$\rep{(2,P)}$~~ & ~~$2P$~~
\\
~~$2$~~ & ~~$SU(2)^2$~~ & ~~$U(P)$~~ & ~~$(\rep{2,1},\brep{P})^{-x}+(\rep{1,2},\brep{P})^{-y}$~~ & ~~$\rep{(2,1,P)}^{x}+\rep{(1,2,P)}^{y}$~~ & ~~$4P$~~
\\
~~$3$~~ & ~~$Sp(2)$~~ & ~~$Sp(P)$~~ & ~~$\rep{(4,2P)}$~~ & ~~$\rep{(4,2P)}$~~ & ~~$8P$~~
\\
~~$4$~~ & ~~$SU(4)$~~ & ~~$Sp(P)\times Sp(\dot{P})$~~ & ~~$(\rep{4},\rep{2P})+(\rep{4},\rep{2\dot{P}})$~~ & ~~$(\brep{4},\rep{2P})+(\brep{4},\rep{2\dot{P}})$~~ & ~~$8P+8\dot{P}$~~
\\
~~$5$~~ & ~~$SO(7)$~~ & ~~$Sp(P)$~~ & ~~$\rep{(8,2P)}$~~ & ~~$\rep{(8,2P)}$~~ & ~~$16P$~~
\\
~~$6$~~ & ~~$SO(8)$~~ & ~~$U(P)$~~ & ~~$(\rep{8}_s,\brep{P})+(\rep{8}_c,\brep{P})$~~ & ~~$(\rep{8}_s,\rep{P})+(\rep{8}_c,\rep{P})$~~ & ~~$16P$~~
\\
~~$7$~~ & ~~$SO(9)$~~ & ~~$SO(P)$~~ & ~~$(\rep{16},\rep{P})$~~ & ~~$(\rep{16},\rep{P})$~~ & ~~$16P$~~
\\
~~$8$~~ & ~~$SO(10)$~~ & ~~$SO(P)\times SO(\dot{P})$~~ & ~~$(\rep{16},\rep{P},  \rep{1})+(\rep{16},  \rep{1}, \rep{\dot{P}})$~~ & ~~$(\brep{16},\rep{P},  \rep{1})+(\brep{16},  \rep{1}, \rep{\dot{P}})$~~ & ~~$16P+16\dot{P}$~~
\\ \\
\hline
\hline
\end{tabular}
\caption{The groups $SO(q+2)\times S_q(P, \dot{P})$ and their representations $\rep{r}$ for the various allowed values of $(q, P, \dot{P})$. The $x, y$ superscripts refer to the $U(1)$ charges, which we leave unfixed.}\label{SqPPreps}
\end{table}
\\\\
The content is $\rep{G}_2\oplus(1+q+2+r)\rep{V}_2$, where $r$ is fixed by $P, \dot{P}$ as in  \autoref{SqPPreps}. Under the maximal reductive global compact symmetry group
\be
U(1)^{st}\times H \times SU(2) =U(1)^{st}\times SO(q+2)\times S_{q}(P,\dot{P})\times SU(2)\times U(1),
\ee
where $U(1)^{st}$ is the spacetime little group,
the content carries the representations:
\begin{align}\label{nonsymvect}
&\rep{(1,1,1)}_{-4}^0+\rep{(1,1,1)}_{4}^0\\
&\rep{(1,1,2)}_{-3}^1+\rep{(1,1,2)}_{3}^{-1}\nonumber\\
&\rep{(1,1,1)}_{-2}^2+\rep{(1,1,1)}_{2}^{-2}&&+&&\rep{(1,1,1)}_{-2}^{-2}+\rep{(1,1,1)}_{2}^{2}&&+&&\rep{(q+2,1,1)}_{-2}^0+\rep{(q+2,1,1)}_{2}^0&&+&&(\rep{r,1})_{-2}^{-1}+( \brep{r},\rep{1})_{2}^{1}\nonumber\\
&&&&&\rep{(1,1,2)}_{-1}^{-1}+\rep{(1,1,2)}_{1}^{1}&&+&&\rep{(q+2,1,2)}_{-1}^1+\rep{(q+2,1,2)}_{1}^{-1}&&+&&(\rep{r},\rep{2})_{-1}^{0}+( \brep{r},\rep{2})_{1}^{0}\nonumber\\
&&&&&\rep{(1,1,1)}_{0}^{0}+\rep{(1,1,1)}_{0}^{0}&&+&&\rep{(q+2,1,1)}_{0}^2+\rep{(q+2,1,1)}_{0}^{-2}&&+&&(\rep{r},\rep{1})_{0}^{1}+( \brep{r},\rep{1})_{0}^{-1}.\nonumber
\end{align}
Note that  the $1,q+2$ and $r$ vector multiplets fall  into three distinct sectors with different representation theoretic properties. As we shall see, this observation follows from the squaring construction;  the three sets  come from three different terms appearing in the product of the Left and Right theories. Here, $r = 2^{[(q+1)/2]} \dim \rep{def}$, where the square parentheses denote the integer part. The $SO(q+2)\times S_{q}(P,\dot{P})$ representations $\rep{r}$ are  summarised in \autoref{SqPPreps} for the various values of $(q, P, \dot{P})$. See also Table 3 of \cite{deWit:1992wf}, where for $q=3$ and $5$ the group $S_q(P, \dot{P})$ is also identified with $U(P, \mathds{H})$.
Note, $S_q(P, \dot{P})$ enjoys  mod 8 Bott periodicity in $q$, following the standard $\R, \R\oplus\R, \R, \C, \Q, \Q\oplus\Q, \Q, \C\ldots$ pattern, and  is symmetric in $P$ and $\dot{P}$. For  $S_q(P, \dot{P})\cong U(P)$, the  defining representation together with its conjugate representation, $\rep{P}\oplus\overline{\rep{P}}$, admits both a  symplectic  and a symmetric real quadratic form that should  be used for a pseudo-real or real gauge group representation respectively, where in the latter case the $Sp$ and $SO$ groups of \autoref{SqPPreps} are interchanged.  

\subsubsection{Generic Jordan}\label{genJ}

There are particular choices of $(q,P,\dot{P})$ for which the non-reductive terms of \eqref{SKnon}  are accompanied by their oppositely charged, under the global
$SO(1,1)$, counterparts. As described  in \autoref{prodsec}, see in particular the discussion around \eqref{magenhance}, this implies the so-called \textit{1st enhancement} to $SL(2, \R)$.  In these cases, the resulting coset spaces are symmetric.
The simplest  example occurs for $(q,P,\dot{P})=(q,0,0)$, implying that $r=0$, which yields the generic Jordan series. Alternatively, the generic Jordan series is also given by $(q,P,\dot{P})=(0,P,0)$ or $(0, 0 , \dot{P})$ \cite{deWit:1992wf}. In this case, the scalar coset is:
\be\label{jordan}
\frac{G}{H}\times\frac{SU(2)}{SU(2)}=\frac{SU(1,1)}{U(1)_g}\times\frac{SO(q+2,2)}{SO(q+2)\times U(1)}\times\frac{SU(2)}{SU(2)}.
\ee
The content is $\rep{G}_2\oplus(1+q+2)\rep{V}_2$ and under
\be
U(1)^{st}\times H\times SU(2)=U(1)^{st}\times SO(q+2)\times SU(2)\times U(1)_g\times U(1)
\ee
carries the following representations:
\begin{align}
&\rep{(1,1)}_{-4}^{(0,0)}+\rep{(1,1)}_{4}^{(0,0)}\nonumber\\
&\rep{(1,2)}_{-3}^{(1,1)}+\rep{(1,2)}_{3}^{(-1,-1)}\nonumber\\
&\rep{(1,1)}_{-2}^{(2,2)}+\rep{(1,1)}_{2}^{(-2,-2)}&&+&&\rep{(1,1)}_{-2}^{(2,-2)}+\rep{(1,1)}_{2}^{(-2,2)}&&+&&\rep{(q+2,1)}_{-2}^{(-2,0)}+\rep{(q+2,1)}_{2}^{(2,0)}\nonumber\\
&&&&&\rep{(1,2)}_{-1}^{(3,-1)}+\rep{(1,2)}_{1}^{(-3,1)}&&+&&\rep{(q+2,2)}_{-1}^{(-1,1)}+\rep{(q+2,2)}_{1}^{(1,-1)}\nonumber\\
&&&&&\rep{(1,1)}_{0}^{(4,0)}+\rep{(1,1)}_{0}^{(-4,0)}&&+&&\rep{(q+2,1)}_{0}^{(0,2)}+\rep{(q+2,1)}_{0}^{(0,-2)}.\nonumber
\end{align}
Again, from the content it is obvious that the $1$ and $q+2$ vector multiplets enjoy a different status. As we shall see in \autoref{squares} the vector multiplets of this theory follow from two different  terms appearing in the product of the Left and Right Yang-Mills factors, which implies that this \textit{1st enhancement} will happen before squaring.

\subsubsection{Magic}\label{vectormagic}

A \textit{1st enhancement} analogous to the previous one occurs also for $(q,P,\dot{P})=(n,1,0)$ where $n=\text{dim}\Al=1,2,4,8,$ for $\Al=\R, \C, \Q, \Oct$ respectively, which implies that $r=2n$. In these cases there is an accidental \textit{2nd enhancement} due to the maximal embedding $[Str_0( \mathfrak{J}^{\Al_\mathds{C}}_3)]_c\supset SO(n+2)\times S_n(1,0)\times U(1)''$ such that the resulting coset is:
\be
\frac{G}{H}\times\frac{SU(2)}{SU(2)}=\frac{Conf( \mathfrak{J}^{\Al}_{3})}{[Str_0( \mathfrak{J}^{\Al_\mathds{C}}_3)]_c\times U(1)'}\times\frac{SU(2)}{SU(2)},
\ee
where $\Al_\mathds{C}\cong\mathds{C}\otimes\Al$, $\mathfrak{J}^{\Al}_{3}$ is the cubic Jordan algebra of $3\times3$ Hermitian matrices over $\Al$ and $ \mathfrak{J}^{\Al_\mathds{C}}_3\cong  \mathds{C}\otimes\mathfrak{J}^{\Al}_3$  its complexification, $Conf(\mathfrak{J})$ is the conformal group of the   cubic Jordan algebra $\mathfrak{J}$, $Str_0(\mathfrak{J})$ is the reduced structure group and $[G]_c$ denotes the compact real form of the  complexified group $G$.
The content is $\rep{G}_2\oplus d\rep{V}_2$ where $d=1+n+2+2n=3(n+1)$, which  under
\be
U(1)^{st}\times H\times SU(2)=U(1)^{st}\times [Str_0( \mathfrak{J}^{\Al_\mathds{C}}_3)]_c\times SU(2)\times U(1)'
\ee
transforms as
\begin{align}
&\rep{(1,1)}_{-4}^{0}+\rep{(1,1)}_{4}^{0}\nonumber\\
&\rep{(1,2)}_{-3}^{3}+\rep{(1,2)}_{3}^{-3}\nonumber\\
&\rep{(1,1)}_{-2}^{6}+\rep{(1,1)}_{2}^{-6}&&+&&(\brep{d},\rep{1})_{-2}^{-2}+(\rep{d},\rep{1})_{2}^{2}\nonumber\\
&&&&&(\brep{d},\rep{2})_{-1}^{1}+(\rep{d},\rep{2})_{1}^{-1}\nonumber\\
&&&&&(\brep{d},\rep{1})_{0}^{4}+(\rep{d},\rep{1})_{0}^{-4},
\end{align}
where the representations $\rep{d}$ are given in \autoref{Jreps}.\\
\begin{table}[h!]
\begin{tabular}{C{1cm} C{2.5cm} C{2.5cm} C{1.5cm} | C{4cm} C{3cm}}
\hline\hline
&&&&& \\
~~$n$~~ & ~~$Conf(\mathfrak{J}^{\Al}_{3})$~~ & ~~$[Str_0(\mathfrak{J}^{\Al_\mathds{C}}_{3})]_c$~~ & ~~$\rep{d}$~~ & ~~$SO(n+2)\times S_n(1,0)$~~ & ~~$\rep{r}$~~
\\
&&&&& \\
\hline
&&&&& \\
~~$1$~~ & ~~$Sp(6;\R)$~~ & ~~$SU(3)$~~ & ~~$\rep{6}$~~ & ~~$SU(2)$~~ & ~~$\rep{2}$~~
\\
~~$2$~~ & ~~$SU(3,3)$~~ & ~~$SU(3)^2$~~ & ~~$\rep{(3,3)}$~~ & ~~$SU(2)^2\times U(1)$~~ & ~~$(\rep{2,1})^{-x}+(\rep{1,2})^{-y}$~~
\\
~~$4$~~ & ~~$SO^\star(12)$~~ & ~~$SU(6)$~~ & ~~$\rep{15}$~~ & ~~$SU(4)\times Sp(1)$~~ & ~~$(\rep{4},\rep{2})$~~
\\
~~$8$~~ & ~~$E_{7(-25)}$~~ & ~~$E_6$~~ & ~~$\rep{27}$~~ & ~~$SO(10)$~~ & ~~$\rep{16}$~~
\\
&&&&& \\
\hline
\hline
\end{tabular}
\caption{Groups and representations appearing the in the 2nd enhancement for the magic supergravities.}\label{Jreps}
\end{table}
\\
The \textit{2nd enhancement} is possible because in all four cases there is a representation $\rep{d}$, which under $[Str_0( \mathfrak{J}^{\Al_\mathds{C}}_3)]_c\supset SO(n+2)\times S_n(1,0)\times U(1)''$  branches to:
\be
\rep{d}\rightarrow\rep{(1,1)}^{-4}+\rep{(q+2,1)}^2+\rep{r}^{-1},
\ee
where the groups and corresponding representations are given in   \autoref{Jreps}.\\
\\
Note, there is a unified description of the 2nd enhancement for magic theories:
they lie in the ``complexified projective planes''
$(\mathds{C} \otimes \mathds{A})\mathds{P}^2$. Namely, the enhancement terms lie in the compact symmetric coset
\be\label{magenhance2}
\frac{[Str_0( \mathfrak{J}^{\Al_\mathds{C}}_3)]_c }{
[Str_0(\mathds{C}\oplus \mathfrak{J}^{\Al_\mathds{C}}_{2})]_c  \times S_n(1,0)},
\ee
where
$[Str_0(\mathds{C}\oplus \mathfrak{J}^{\Al_\mathds{C}}_{2})]_c   \times  S_n(1,0)= SO(q+2) \times  S_n(1,0) \times U(1)''$ and $S_n(1,0) = S_q(1, 0)$. Note that $S_q(1, 0) = S_q(0, 1) = \text{Id}, U(1), Sp(1), \text{Id}$ for $q=1,2,4,8$ (or, equivalently, $S_q(1, 0) = S_q(0, 1) =tri(\Al)/so(\Al)$, where $tri(\Al)$ and $so(\Al)$ respectively denote the triality and orthogonal symmetries of $\Al=\R, \C, \Q, \Oct$ \cite{Baez:2001dm}) and the symmetric embedding $[Str_0(\mathds{C}\oplus \mathfrak{J}^{\Al_\mathds{C}}_{2})]_c  \times S_n(1,0)\subset  [Str_0( \mathfrak{J}^{\Al_\mathds{C}}_3)]_c$ follows from  the maximal Jordan algebra  embedding,
$\mathds{C}\oplus \mathfrak{J}^{\Al_\mathds{C}}_{2} \subset \mathfrak{J}^{\Al_\mathds{C}}_3$.
The tangent space of \eqref{magenhance2}  can be represented  as $\Al_\mathds{C}\oplus\Al_\mathds{C}$, where the summands carry equal and opposite $U(1)''$ charges (specifically 3 vs. $-3$). Namely, they are a pair of chiral spinors in $D=q+2$ critical dimensions; for $q=1$ there is no chiral splitting. Note, these groups are the compact analogs of the $SO(1,q+1) \times S_n(1,0)$
 U-duality groups of the corresponding $D=6$  magic supergravity theories and the $U(1)''$ appearing in the stabilizer of \eqref{magenhance2} is the compact version of the Kaluza-Klein $SO(1,1)$ of the compactification from $D=6$ to   $D=5$.\\
\\
The vector multiplets $1,q+2$ and $r$ come from three different terms appearing in the product of the Left and Right Yang-Mills factors, which implies that the 2nd enhancement  only appears after squaring. Hence, for the purpose of matching the content to that obtained  from the Left and Right Yang-Mills factors, we should first  decompose under $U(1)^{st}\times SO(n+2)\times S_n(1,0)\times SU(2)\times U(1)'\times U(1)''$:
\begin{align}
&\rep{(1,1,1)}_{-4}^{(0,0)}+\rep{(1,1,1)}_{4}^{(0,0)}\\
&\rep{(1,1,2)}_{-3}^{(3,0)}+\rep{(1,1,2)}_{3}^{(-3,0)}\nonumber\\
&\rep{(1,1,1)}_{-2}^{(6,0)}+\rep{(1,1,1)}_{2}^{(-6,0)}&&+&&\rep{(1,1,1)}_{-2}^{(-2,4)}+\rep{(1,1,1)}_{2}^{(2,-4)}&&+&&\rep{(q+2,1,1)}_{-2}^{(-2,-2)}+\rep{(q+2,1,1)}_{2}^{(2,2)}\nonumber\\
&&&+&&(\rep{r},\rep{1})_{-2}^{(-2,1)}+(\brep{r},\rep{1})_{2}^{(2,-1)}\nonumber\\
&&&&&\rep{(1,1,2)}_{-1}^{(1,4)}+\rep{(1,1,2)}_{1}^{(-1,-4)}&&+&&\rep{(q+2,1,2)}_{-1}^{(1,-2)}+\rep{(q+2,1,2)}_{1}^{(-1,2)}\nonumber\\
&&&+&&(\rep{r},\rep{2})_{-1}^{(1,1)}+(\brep{r},\rep{2})_{1}^{(-1,-1)}\nonumber\\
&&&&&\rep{(1,1,1)}_{0}^{(4,4)}+\rep{(1,1,1)}_{0}^{(-4,-4)}&&+&&\rep{(q+2,1,1)}_{0}^{(4,-2)}+\rep{(q+2,1,1)}_{0}^{(-4,2)}\nonumber\\
&&&+&&(\rep{r},\rep{1})_{0}^{(4,1)}+(\brep{r},\rep{1})_{0}^{(-4,-1)}.\nonumber
\end{align}

\subsubsection{Minimally coupled}\label{mcoup}

The three classes of theories summarised above relied on the method of classifying SK manifolds through their dimensional reduction from $D=5$ and are thus  unified in their description. The minimally coupled supergravities, however, do not in general admit an oxidation to $D=5$ and  so stand on their own,  not fitting the general analysis of the previous three classes. Note, however, the Yang-Mills squared description unifies  them all, as we shall see. The minimally coupled symmetric scalar coset is:
\be\label{min}
\frac{G}{H}\times\frac{SU(2)}{SU(2)}=\frac{SU(1,P+1)}{U(1)_{min}\times SU(P+1)}\times\frac{U(1)}{U(1)}\times\frac{SU(2)}{SU(2)},
\ee
which is isomorphic to the hyperbolic projective space $\overline{\mathds{CP}}^{P+1}$. For more details, see Table 9 of \cite{deWit:1995tf} and the associated comments. Their content is given by $\rep{G}_2\oplus(P+1)\rep{V}_2$, which under
\be
U(1)^{st}\times H\times SU(2)=U(1)^{st}\times SU(P+1)\times SU(2)\times U(1)_{min}\times U(1)
\ee
transforms as
\begin{align}\label{mcreps}
&\rep{(1,1)}_{-4}^{(0,0)}+\rep{(1,1)}_{4}^{(0,0)}&&\\
&\rep{(1,2)}_{-3}^{(P+1,1)}+\rep{(1,2)}_{3}^{(-P-1,-1)}&&\nonumber\\
&\rep{(1,1)}_{-2}^{(2P+2,2)}+\rep{(1,1)}_{2}^{(-2P-2,-2)}&&+&&\rep{(P+1,1)}_{-2}^{(2,-2)}+(\brep{P+1},\rep{1})_{2}^{(-2,2)}&&\nonumber\\
&&&&&\rep{(P+1,2)}_{-1}^{(P+3,-1)}+(\brep{P+1},\rep{2})_{1}^{(-3-3,1)}&&\nonumber\\
&&&&&\rep{(P+1,1)}_{0}^{(2P+4,0)}+(\brep{P+1},\rep{1})_{0}^{(-2P-4,0)}.&&\nonumber
\end{align}
Anticipating the Yang-Mills squared construction, the $1$ and $P$ vector multiplets will necessarily come from two distinct   terms in the product. Consequently, in order to match the content with that obtained by squaring we need to further decompose  $SU(P+1)\supset SU(P)\times U(1)_P$. Under the resulting 
\be
U(1)^{st}\times SU(P)\times SU(2)\times U(1)_{min}\times U(1)\times U(1)_P
\ee
the content  transforms as:
 \begin{align}
&\rep{(1,1)}_{-4}^{(0,0,0)}+\rep{(1,1)}_{4}^{(0,0,0)}&&\\
&\rep{(1,2)}_{-3}^{(P+1,1,0)}+\rep{(1,2)}_{3}^{(-P-1,-1,0)}&&\nonumber\\
&\rep{(1,1)}_{-2}^{(2P+2,2,0)}+\rep{(1,1)}_{2}^{(-2P-2,-2,0)}&&+&&\rep{(1,1)}_{-2}^{(2,-2,-P)}+(\rep{1},\rep{1})_{2}^{(-2,2,P)}&&+&&\rep{(P,1)}_{-2}^{(2,-2,1)}+(\brep{P},\rep{1})_{2}^{(-2,2,-1)}&&\nonumber\\
&&&&&\rep{(1,2)}_{-1}^{(P+3,-1,-P)}+(\rep{1},\rep{2})_{1}^{(-P-3,1,P)}&&+&&\rep{(P,2)}_{-1}^{(P+3,-1,1)}+(\brep{P},\rep{2})_{1}^{(-P-3,1,-1)}&&\nonumber\\
&&&&&\rep{(1,1)}_{0}^{(2P+4,0,-P)}+(\rep{1},\rep{1})_{0}^{(-2P-4,0,P)}&&+&&\rep{(P,1)}_{0}^{(2P+4,0,1)}+(\brep{P},\rep{1})_{0}^{(-2P-4,0,-1)}.&&\nonumber
\end{align}
Naively, this can be regarded as the extension of \autoref{SqPPreps} to $(q,P,\dot{P})=(-2,P,0)$ with $\rep{r}=\rep{P}^{-1}+\brep{P}^1$ of $U(P)$ \cite{deWit:1995tf}.

\subsubsection{$T^3$ model}

Like the generic Jordan and magic supergravities, the $T^3$ model can be constructed using a cubic Jordan algebra, namely $\mathfrak{J}_3\cong\mathds{R}$. However, it does not strictly sit in the  generic Jordan sequence, rather it should be considered as the ``symmetrization'' of the $q=0$ generic Jordan supergravity (otherwise known as the $STU$ model \cite{Duff:1995sm}). This is reflected by the fact that it follows from the dimensional reduction of ``pure''  $\mathcal{N}=2, D=5$ supergravity. The $T^3$ scalar coset is:
\be\label{t3coset}
\frac{G}{H}\times\frac{SU(2)}{SU(2)}=\frac{SU(1,1)}{U(1)_T}\times\frac{SU(2)}{SU(2)}.
\ee
The content $\rep{G}_2\oplus \rep{V}_2$ transforms under
\be
U(1)^{st}\times H\times 
SU(2)=U(1)^{st}\times SU(2)\times U(1)_T
\ee
as
\begin{align}\label{T3reps}
&\rep{1}_{-4}^{0}+\rep{1}_{4}^{0}&&\\
&\rep{2}_{-3}^{1}+\rep{2}_{3}^{-1}&&\nonumber\\
&\rep{1}_{-2}^{2}+\rep{1}_{2}^{-2}&&+&&\rep{1}_{-2}^{-6}+\rep{1}_{2}^{6}&&\\
&&&&&\rep{2}_{-1}^{-5}+\rep{2}_{1}^{5}&&\nonumber\\
&&&&&\rep{1}_{0}^{-4}+\rep{1}_{0}^{4}.&&\nonumber
\end{align}
It is not possible to generate the $T^3$ model from Yang-Mills squared, at least not straightforwardly, although the  content and representations presented above \emph{can} be reproduced by the product of $\mathcal{N}_L=2$ and $\mathcal{N}_R=0$ Yang-Mills theories. We will address these two comments properly in  \autoref{vectors}.
\subsection{Hyper multiplets}\label{hyper}

Here we consider the inclusion of hyper multiplets. The coset manifold parametrised by the hyper scalars must be  quaternionic  \cite{Cecotti:1988qn, Cecotti:1988ad, deWit:1992wf}. In the previous subsection the   $SU(2)$ R-symmetry  was included as an additional factor commuting  with isometries of the   SK scalar manifolds. When  hyper multiplets are included this $SU(2)$ factor is absorbed by the Q manifold of the hyper scalars. Specifically, it becomes part of the Q manifold holonomy group, which for homogeneous manifolds is in turn part of the isotropy group.

\subsubsection{Non-symmetric}\label{hnonsym}

The non-symmetric  hyper scalar Q manifold is the c-map \cite{deWit:1995tf} of the non-symmetric SK manifold given in \eqref{SKnon},
\be\label{hypcoset}
\frac{\mathcal{G}}{\mathcal{H}} =SO(1,1)\times\frac{SO(q+3,3)}{SO(q+3)\times SU(2)_{ns}}\times\frac{S_{q}(P,\dot{P})}{S_{q}(P,\dot{P})}\ltimes\Big[(\rep{spin},\rep{def})^1\ltimes(\rep{q+6,1})^2\Big]
\ee
Note, the  $(q, P, \dot{P})$ used here are independent of those appearing   in the SK manifold \eqref{SKnon}. The additional content is $(q+4+s/2)\rep{H}_2$, which under
\be
U(1)^{st}\times \mathcal{H}=U(1)^{st}\times SO(q+3)\times S_q(P,\dot{P})\times SU(2)_{ns}.
\ee
transforms as
\begin{align}\label{hypref}
&(\rep{q+3,1,2})_{-1}+(\rep{q+3,1,2})_{1}&&+&&(\rep{1,1,2})_{-1}+(\rep{1,1,2})_{1}&&+&&(\rep{s,1})_{-1}+(\rep{s,1})_{1}\\\nonumber
&(\rep{q+3,1,3})_0+(\rep{q+3,1,1})_0&&+&&(\rep{1,1,3})_0+(\rep{1,1,1})_0&&+&&(\rep{s,2})_0,
\end{align}
where $\rep{s}$ is a not necessarily  irreducible representation of  $SO(q+3)\times S_q(P,\dot{P})$, as given in \autoref{Hyprep}, of dimension
$s= 2^{[(q+3)/2]} \dim \rep{def}$.

\begin{table}[h!]
\begin{tabular}{C{1cm} C{3cm} C{3cm} C{6cm} C{2cm}}
\hline\hline
\\
~~$q$~~ & ~~$SO(q+3)$~~ & ~~$S^q(P,\dot{P})$~~ & ~~$\rep{s}$~~ & ~~$s(q,P,\dot{P})$~~
\\ \\
\hline \\
~~$-2$~~ & ~~-//-~~ & ~~$U(P)$~~ & ~~$\rep{P}^{-x}+\brep{P}^{x}$~~ & ~~$2P$~~
\\
~~$-1$~~ & ~~$U(1)$~~ & ~~$SO(P)$~~ & ~~$\rep{P}^{-x}+\rep{P}^{x}$~~ & ~~$2P$~~
\\
~~$0$~~ & ~~$SU(2)$~~ & ~~$SO(P)\times SO(\dot{P})$~~ & ~~$\rep{(2,P)}+\rep{(2,\dot{P})}$~~ & ~~$2P+2\dot{P}$~~
\\
~~$1$~~ & ~~$SU(2)^2$~~ & ~~$SO(P)$~~ & ~~$\rep{(2,1,P)}+\rep{(1,2,P)}$~~ & ~~$4P$~~
\\
~~$2$~~ & ~~$Sp(2)$~~ & ~~$U(P)$~~ & ~~$(\rep{4},\brep{P})^{-x}+\rep{(4,P)}^{x}$~~ & ~~$8P$~~
\\
~~$3$~~ & ~~$SU(4)$~~ & ~~$Sp(P)$~~ & ~~$\rep{(4,2P)}+(\brep{4},\rep{2P)}$~~ & ~~$16P$~~
\\
~~$4$~~ & ~~$SO(7)$~~ & ~~$Sp(P)\times Sp(\dot{P})$~~ & ~~$(\rep{8},\rep{2P})+(\rep{8},\rep{2\dot{P}})$~~ & ~~$16P+16\dot{P}$~~
\\
~~$5$~~ & ~~$SO(8)$~~ & ~~$Sp(P)$~~ & ~~$(\rep{8}_s,\rep{2P})+(\rep{8}_c,\rep{2P})$~~ & ~~$32P$~~
\\
~~$6$~~ & ~~$SO(9)$~~ & ~~$U(P)$~~ & ~~$(\rep{16},\rep{P})^{-x}+(\rep{16},\brep{P})^{x}$~~ & ~~$32P$~~
\\
~~$7$~~ & ~~$SO(10)$~~ & ~~$SO(P)$~~ & ~~$(\rep{16},\rep{P})+(\brep{16},\rep{P})$~~ & ~~$32P$~~
\\
~~$8$~~ & ~~$SO(11)$~~ & ~~$SO(P)\times SO(\dot{P})$~~ & ~~$(\rep{32},\rep{P})+(\rep{32},\rep{\dot{P}})$~~ & ~~$32P+32\dot{P}$~~
\\ \\
\hline
\hline
\end{tabular}
\caption{Groups and representations appearing in \eqref{hypcoset} and \eqref{hypref}.  Note, the $q=-2$ case  corresponds to the c-map of the minimally coupled vector multiplet series \eqref{min}, see  Table 9 of  \cite{deWit:1995tf}. Again, $S_q(P, \dot{P})$ enjoys the standard mod 8 Bott periodic pattern in $q$.}\label{Hyprep}
\end{table}

\subsubsection{Generic Jordan}\label{hgenJ}

As for the vector multiplet sector, there are  particular choices of $(q,P,\dot{P})$ for which the non-reductive terms appearing in \eqref{hypcoset}  carry representations that allow for the so-called
\textit{1st enhancement} $SO(1,1)\times SO(q+3, 3)\subset SO(q+4,4)$.  In this case the representations $\rep{s}$ enhance to $\rep{t}$ as given in \autoref{stot}, and the resulting coset spaces are symmetric.
\begin{table}[h!]
\begin{tabular}{C{1cm} C{3cm} C{3cm} C{6cm} C{2cm}}
\hline\hline
\\
~~$q$~~ & ~~$SO(q+4)$~~ & ~~$S_q(P,\dot{P})$~~ & ~~$\rep{t}$~~ & ~~$t(q,P,\dot{P})$~~
\\ \\
\hline \\
~~$-3$~~ & ~~-//-~~ & ~~$Sp(P)$~~ & ~~$\rep{2P}$~~ & ~~$2P$~~
\\
~~$-2$~~ & ~~$U(1)$~~ & ~~$U(P)$~~ & ~~$\rep{P}^{(-x,a)}+\brep{P}^{(x,-a)}$~~ & ~~$2P$~~
\\
~~$-1$~~ & ~~$SU(2)$~~ & ~~$SO(P)$~~ & ~~$\rep{(2,P)}$~~ & ~~$2P$~~
\\
~~$0$~~ & ~~$SU(2)^2$~~ & ~~$SO(P)\times SO(\dot{P})$~~ & ~~$\rep{(2,1,P, 1)}+\rep{(1,2,1, \dot{P})}$~~ & ~~$2P+2\dot{P}$~~
\\
~~$1$~~ & ~~$Sp(2)$~~ & ~~$SO(P)$~~ & ~~$\rep{(4,P)}$~~ & ~~$4P$~~
\\
~~$2$~~ & ~~$SU(4)$~~ & ~~$U(P)$~~ & ~~$(\rep{4},\brep{P})^{-x}+(\brep{4},\rep{P)}^{x}$~~ & ~~$8P$~~
\\
~~$3$~~ & ~~$SO(7)$~~ & ~~$Sp(P)$~~ & ~~$\rep{(8,2P)}$~~ & ~~$16P$~~
\\
~~$4$~~ & ~~$SO(8)$~~ & ~~$Sp(P)\times Sp(\dot{P})$~~ & ~~$(\rep{8}_s,\rep{2P})+(\rep{8}_c,\rep{2\dot{P}})$~~ & ~~$16P+16\dot{P}$~~
\\
~~$5$~~ & ~~$SO(9)$~~ & ~~$Sp(P)$~~ & ~~$(\rep{16},\rep{2P})$~~ & ~~$32P$~~
\\
~~$6$~~ & ~~$SO(10)$~~ & ~~$U(P)$~~ & ~~$(\rep{16},\brep{P})^{-x}+(\brep{16},\rep{P})^x$~~ & ~~$32P$~~
\\
~~$7$~~ & ~~$SO(11)$~~ & ~~$SO(P)$~~ & ~~$(\rep{32},\rep{P})$~~ & ~~$32P$~~
\\
~~$8$~~ & ~~$SO(12)$~~ & ~~$SO(P)\times SO(\dot{P})$~~ & ~~$(\rep{32}, \rep{P}, \rep{1})+(\rep{32}, \rep{1}, \rep{\dot{P}})$~~ & ~~$32P+32\dot{P}$~~
\\ \\
\hline
\hline
\end{tabular}
\caption{Groups and representations for hyper scalar Q manifolds with the 1st enhancement $\rep{s}\rightarrow\rep{t}$. The case $q = -3$ is given in Table 9 of \cite{deWit:1995tf} and it is nothing but the series of hyperbolic quaternionic projective spaces  $\overline{\Q\mathds{P}}^{P+1}$, which are not in the c-map image of any (projective) special K\"ahler manifold.}\label{stot}
\end{table}
The simplest  example is the c-map of the generic Jordan series \eqref{jordan}, which occurs for $(q,P,\dot{P})=(q,0,0)$,  implying that $s=0$, or alternatively, for  $(q,P,\dot{P})=(0,P,0)$ or $(0, 0 , \dot{P})$ \cite{deWit:1992wf}.  The resulting symmetric hyper scalar coset is given by,
\be\label{hypJcoset}
\frac{\mathcal{G}}{\mathcal{H}} =\frac{SO(q+4,4)}{SO(q+4)\times SU(2)\times SU(2)'}.
\ee
The additional content is $(q+4)\rep{H}_2$ which under
\be
U(1)^{st}\times \mathcal{H}=U(1)^{st}\times SO(q+4)\times SU(2)\times SU(2)'
\ee
transforms as,
\be
\rep{(q+4,1,2})_{-1}+(\rep{q+4,2,2})_0+ \rep{(q+4,1,2})_{1}.
\ee
As we shall see in \autoref{squares}  this \textit{1st enhancement} will happen before squaring, in the sense that the Right Yang-Mills theory itself has a global $SO(q+4)$ symmetry that becomes the $SO(q+4)$ of \eqref{hypJcoset}.

\subsubsection{Magic}\label{hvectormagic}

A \textit{1st enhancement} analogous to the previous one occurs also for $(q,P,\dot{P})=(n,1,0)$ where $n=\text{dim}\Al=1,2,4,8$ which implies that $t=4n$. In these cases there is an accidental \textit{2nd enhancement} due to the maximal embedding $[Conf(\mathfrak{J}^{\Al_\C}_{3})]_c\supset SO(n+4)\times S_n(1,0)\times SU(2)'$ such that the resulting coset is
\be
\frac{\mathcal{G}}{\mathcal{H}} =\frac{QConf(\mathfrak{J}^{\Al}_{3})}{[Conf(\mathfrak{J}^{\Al_\C}_{3})]_c\times SU(2)},
\ee
where $QConf(\mathfrak{J})$ is the quasi-conformal group of the   Jordan algebra $\mathfrak{J}$. The additional content is $(f/2)\rep{H}_2$ where $f=2(3n+4)$, which under
\be
U(1)^{st}\times \mathcal{H}=U(1)^{st}\times [Conf(\mathfrak{J}^{\Al_\C}_{3})]_c\times SU(2)
\ee
transforms as
\be\label{magicbreak1}
(\rep{f,1})_1+(\rep{f,2})_0+(\rep{f,1})_{-1}.
\ee
The \textit{2nd enhancement} is possible because in all four cases there is a symplectic representation $\rep{f}$, which under $[Conf(\mathfrak{J}^{\Al_\C}_{3})]_c\supset SO(n+4)\times S_n(1,0)\times SU(2)'$ decomposes as,
\be
\rep{f}\rightarrow\rep{(q+4,1,2)+(t,1)},
\ee
where $\rep{f}$ and $\rep{t}$ are given in \autoref{maghyptab}.
\begin{table}[h!]
\begin{tabular}{C{1cm} C{2.5cm} C{2.5cm} C{1.5cm} | C{4cm} C{3cm}}
\hline\hline
&&&&& \\
~~$n$~~ & ~~$QConf(J^\Al_3)$~~ & ~~$[Conf(\mathfrak{J}^{\Al_\C}_{3})]_c$~~ & ~~$\rep{f}$~~ & ~~$SO(n+4)\times S_n(1,0)$~~ & ~~$\rep{t}$~~
\\
&&&&& \\
\hline
&&&&& \\
~~$1$~~ & ~~$F_{(4(4))}$~~ & ~~$Sp(3)$~~ & ~~$\rep{14'}$~~ & ~~$Sp(2)$~~ & ~~$\rep{4}$~~
\\
~~$2$~~ & ~~$E_{6(2)}$~~ & ~~$SU(6)$~~ & ~~$\rep{20}$~~ & ~~$SU(4)\times U(1)$~~ & ~~$\rep{4}^{-x}+\brep{4}^{x}$~~
\\
~~$4$~~ & ~~$E_{7(-5)}$~~ & ~~$SO(12)$~~ & ~~$\rep{32}$~~ & ~~$SO(8)\times Sp(1)$~~ & ~~$(\rep{8}_s,\rep{2})$~~or~~$(\rep{8}_c,\rep{2})$ ~~
\\
~~$8$~~ & ~~$E_{8(-24)}$~~ & ~~$E_7$~~ & ~~$\rep{56}$~~ & ~~$SO(12)$~~ & ~~$\rep{32}$~~
\\
&&&&& \\
\hline
\hline
\end{tabular}
\caption{The $\mathcal{H}$ representations $\rep{f}$ carried by the magic hyper scalars and the representations $\rep{t}$ appearing in the breaking $\rep{f}\rightarrow\rep{(q+4,1,2)+(t,1)}$ under $SO(n+4)\times S_n(1,0)\times SU(2)'\subset [Conf(\mathfrak{J}^{\Al_\C}_{3})]_c$. 
 }\label{maghyptab}
\end{table}
As we will see in \autoref{squares} the hyper multiplets  come exclusively  from the product of Left factor half-hyper multiplets with Right factor scalars, so these enhancements happen before squaring.\\
\\
Note, as for the magic vector multiplet case given in \autoref{vectormagic}, there is a unified description of the 2nd enhancement for magic theories:
they lie in the ``quaternionified  projective planes''
$(\mathds{H} \otimes \mathds{A})\mathds{P}^2$. Namely, the enhancement terms lie in the compact symmetric coset
\be\label{QP2}
[Conf( \mathfrak{J}^{\Al_\mathds{C}}_3)]_c /
[Conf(\mathds{C}\oplus \mathfrak{J}^{\Al_\mathds{C}}_{2})]_c  \times S_n(1,0),
\ee
where
$[Conf(\mathds{C}\oplus \mathfrak{J}^{\Al_\mathds{C}}_{2})]_c   \times  S_n(1,0)= SO(q+4) \times SU(2)' \times  S_n(1,0) $. The symmetric embedding $[Conf(\mathds{C}\oplus \mathfrak{J}^{\Al_\mathds{C}}_{2})]_c  \times S_n(1,0)\subset  [Conf( \mathfrak{J}^{\Al_\mathds{C}}_3)]_c$ follows from  the maximal Jordan algebra  embedding,
$\mathds{C}\oplus \mathfrak{J}^{\Al_\mathds{C}}_{2} \subset \mathfrak{J}^{\Al_\mathds{C}}_3$.
The tangent space of \eqref{QP2}  can be represented  as $\Q\otimes\Al\oplus\Q\otimes\Al$, where the summands transform as a doublet under  $SU(2)'$. Namely, they are a pair of chiral spinors in $D=q+4$  dimensions; for $q=1$ there is no chiral splitting.

\subsubsection{Minimally coupled}\label{hmcoup}

For the above cases the quaternionic manifolds of the hyper sector are insensitive to dimensional reduction, and therefore are the same in $D=3,4,5,6$. These spaces can be constructed by composing the r-map and c-map into a map (termed the q-map in mathematical literature, cfr.~e.g.~\cite{Cortes:2017yrp}) from real special homogeneous manifolds.  By contrast, for the addition of hyper multiplets to the minimally coupled model of \autoref{mcoup} there is no real special homogeneous starting point (i.e.~there is no r-map and hence no q-map). However  the scalar coset,
\be
\frac{\mathcal{G}}{\mathcal{H}} =\frac{SU(P,2)}{U(P)\times SU(2)},
\ee
 is given by the c-map of the SK minimally coupled series \eqref{min}, as one would anticipate. The additional content is $P\rep{H}_2$, which under
 \be
 U(1)^{st}\times \mathcal{H}=U(1)^{st}\times U(P)\times SU(2)
 \ee
transforms as
\be
(\rep{P},\rep{1})_{-1}^{-1}+(\brep{P},\rep{1})_{-1}^{1}+(\rep{P},\rep{2})_{0}^{-1}+(\brep{P},\rep{2})_{0}^{1}+(\rep{P},\rep{1})_{1}^{-1}+(\brep{P},\rep{1})_{1}^{1}.
\ee
This can be regarded as the $(q,P,\dot{P})=(-4,P,0)$ extension of \autoref{stot}, with $\rep{t}=\rep{P}^{-1}+\brep{P}^{1}$ of $U(P)$,  as given in Table 9 of  \cite{deWit:1995tf}.

\subsubsection{Projective quaternionic}\label{hproj}

A second possibility for  minimally coupled hyper multiplets  is given by
\be
\frac{\mathcal{G}}{\mathcal{H}} =\frac{Sp(P,1)}{Sp(P)\times SU(2)}.
\ee
These are projective quaternionic symmetric spaces given by $q=-3$ of \autoref{stot} (also cfr.~table 9 of  \cite{deWit:1995tf}). Note, they are not in the c-map image of any (projective) special K\"ahler manifold and in this sense they are distinguished. The additional content is $P\rep{H}_2$, which under
\be
U(1)^{st}\times \mathcal{H}=U(1)^{st}\times Sp(P)\times SU(2)
\ee
transforms as
\be
\rep{(2P,1)}_{-1}+\rep{(2P,2)}_0+\rep{(2P,1)}_{1}.
\ee
This can be regarded as the  $(q,P,\dot{P})=(-3,P,0)$ entry of  \autoref{stot} with $\rep{t}=\rep{2P}$ of $Sp(P)$.

\subsubsection{Exceptional $T^3$ model}\label{hT3}

The final case is given by the inclusion of hyper multiplets in the $T^3$ model. The coset is exceptional,
\be\label{g2}
\frac{\mathcal{G}}{\mathcal{H}} =\frac{G_{2(2)}}{SU(2)_E\times SU(2)},
\ee
and  is the c-map of \eqref{t3coset} or the q-map of a point, reflecting the fact that  dimensionally reducing pure $D=5$ supergravity to $D=3$ yields a scalar coset given by \eqref{g2} once the 1-form potentials have been dualised to scalars. The additional content is $2\rep{H}_2$, which under $U(1)^{st}\times \mathcal{H}=U(1)^{st}\times SU(2)_E\times SU(2)$ transforms as,
\be
\rep{(4,1)}_{-1}+\rep{(4,2)}_0+\rep{(4,1)}_{1}.
\ee

\section{Squaring}\label{squares}

\subsection{General principles}\label{genpric}
The field content generated by all products  of Left and Right multiplets (excluding those generating gravitino multiplets) are given in \autoref{prods}.  These are deduced using the tensor product of asymptotic on-shell helicity states, which we denote by $\otimes$. Of course, this does not fix the couplings or symmetries of the corresponding theory (unless they are  fixed by supersymmetry). However,   following \cite{Anastasiou:2014qba} we can use the convolutive tensor product, $\circ$, of Left and Right spacetime  fields   to deduce the possible symmetries  and hence couplings of the resulting  theory. For Left and Right multiplets $\mathbf{L}, \mathbf{R}$, the product is defined by
\be\label{offshell}
\mathbf{L}\circ\mathbf{R} :=  \mathbf{L}^{\Sigma} \star \Phi_{\Sigma\tilde{\Sigma}} \star \mathbf{R}^{\tilde{\Sigma}},
\ee
where
\be
[f\star g](x)=\int d^Dy f(y) g(x-y)
\ee
for arbitrary spacetime fields $f, g$. The convolution reflects the fact that the amplitude relations are multiplicative in momentum space. It turns out to be essential for reproducing the local symmetries of (super)gravity from those of the two (super) Yang-Mills factors to linear order \cite{Anastasiou:2014qba,  Borsten:2015pla, Anastasiou:2016csv}. The spectator field $\Phi_{\Sigma\tilde{\Sigma}}$ allows for arbitrary and independent   $G_L$ and $ G_R$ at the level of spacetime fields. The indices $\Sigma, \tilde{\Sigma}$ run over the representations carried by the Left and Right multiplets under the Left and Right gauge groups $G_L$ and $G_R$. The spectator takes a  block-diagonal  form,
\be\label{spec}
\Phi = \begin{pmatrix} \Phi_{A\tilde{A}} & 0\\ 0& \Phi_{a\tilde{a}}\end{pmatrix},
\ee
where $A, \tilde{A}$ are adjoint\footnote{Note, the bi-adjoint scalar field $\Phi_{A\tilde{A}} $  plays a crucial role in the Yang-Mills squared construction of classical (supersymmetric) single- and multi-centre black hole solutions \cite{Cardoso:2016ngt, Cardoso:2016amd} and also appears by very close analogy  in  the non-perturbative double-copy construction of  Kerr-Schild solutions  \cite{Monteiro:2014cda, Luna:2015paa, Luna:2016due, White:2016jzc}, although the precise relationship between the two pictures remains an intriguing open question.} and $a, \tilde{a}$ are ``fundamental''\footnote{Here we use ``fundamental'' to mean any (not necessarily irreducible) representations other than the adjoint.}  indices of $G_L$ and $G_R$, respectively. This enforces the fact that adjoint representations only double-copy with adjoint representations \cite{Johansson:2014zca, Anastasiou:2016csv}. As discussed in detail in \cite{Borsten:2013bp, Anastasiou:2014qba, Anastasiou:2013hba, Anastasiou:2015vba, Borsten:2015pla, Anastasiou:2016csv} this product allows us to reconstruct the symmetries of the resulting supergravity theory (under the assumption that the scalar coset manifold is homogeneous) in terms of its Yang-Mills-matter factors and we apply these same principles in the subsequent analysis. Note, the product of the global symmetries of the two factors yields a subset of the gravitational global symmetries, which are enhanced to the full set of generically non-compact global symmetries as described in \aref{enhancements}.
\begin{table}[h!]
\begin{tabular}{L{2.5cm} L{2cm}}
\hline\hline
\\
~~$\rep{L}\otimes\rep{R}$~~ & ~~Result~~
\\ \\
\hline \\
~~$\rep{C}_1\otimes\lambda$~~ & ~~$\rep{V}_1\oplus\rep{C}_1$~~
\\
~~$\rep{C}_1\otimes\rep{C}_1$~~ & ~~$\rep{V}_2\oplus\rep{H}_2$~~
\\
~~$\rep{C}_2\otimes\lambda$~~ & ~~$\rep{V}_2$~~
\\
~~$\rep{C}_2\otimes\rep{C}_1$~~ & ~~$\rep{V}_3$~~
\\
~~$\rep{C}_2\otimes\rep{H}_2$~~ & ~~$\rep{V}_4$~~
\\
~~$\rep{H}_2\otimes\lambda$~~ & ~~$2\rep{V}_2$~~
\\
~~$\rep{H}_2\otimes\rep{C}_1$~~ & ~~$2\rep{V}_3$~~
\\
~~$\rep{H}_2\otimes\rep{H}_2$~~ & ~~$4\rep{V}_4$~~
\\ \\ \\
\hline
\hline
\end{tabular}
\quad\quad\quad\quad\quad
\begin{tabular}{L{2.5cm} L{2cm}}
\hline\hline
\\
~~$\rep{L}\otimes\rep{R}$~~ & ~~Result~~
\\ \\
\hline \\
~~$\rep{V}_1\otimes A$~~ & ~~$\rep{G}_1\oplus\rep{C}_1$~~
\\
~~$\rep{V}_1\otimes\rep{V}_1$~~ & ~~$\rep{G}_2\oplus\rep{H}_2$~~
\\
~~$\rep{V}_2\otimes A$~~ & ~~$\rep{G}_2\oplus\rep{V}_2$~~
\\
~~$\rep{V}_2\otimes\rep{V}_1$~~ & ~~$\rep{G}_3\oplus\rep{V}_3$~~
\\
~~$\rep{V}_2\otimes\rep{V}_2$~~ & ~~$\rep{G}_4\oplus2\rep{V}_4$~~
\\
~~$\rep{V}_4\otimes A$~~ & ~~$\rep{G}_4$~~
\\
~~$\rep{V}_4\otimes\rep{V}_1$~~ & ~~$\rep{G}_5$~~
\\
~~$\rep{V}_4\otimes\rep{V}_2$~~ & ~~$\rep{G}_6$~~
\\
~~$\rep{V}_4\otimes\rep{V}_4$~~ & ~~$\rep{G}_8$~~
\\ \\
\hline
\hline
\end{tabular}
\caption{The content resulting from the product of the on-shell helicity states of Left and Right multiplets, as summarised in \autoref{smults}.}\label{prods}
\end{table}

\subsection{The gauge theory factors}

Here we summarise the Left and Right Yang-Mills-matter theories used  subsequently to generate the supergravity theories described in the previous sections. 

\subsubsection{The Left $\mathcal{N}_L=2$ gauge theory}

The Left theory consists of one $\mathcal{N}=2$ vector multiplet $\rep{V}_2^A$ and one half-hyper multiplet $\rep{C}_2^a$, which  carries  a  pseudoreal fundamental representation\footnote{We also always require that the adjoint of the gauge group is included in the symmetric tensor product of the pseudoreal fundamental.} of the Left gauge group $G_L$. Note, we use $A$ and $a$ to distinguish fields carrying adjoint and fundamental representations, respectively, of the gauge group. The spacetime, global and gauge symmetries are given by
\be
U(1)^{st}_L\times SU(2)_L\times U(1)_L\times G_L,
\ee
 under which the multiplets transform as follows:
\begin{align}
\rep{V}_2^A&:\Big[\rep{1}_{-2}^0+\rep{1}_{2}^0+\rep{2}_{-1}^1+\rep{2}_{1}^{-1}+\rep{1}_{0}^2+\rep{1}_{0}^{-2}\Big]^A,\\
\rep{C}_2^a&:\Big[\rep{1}_{-1}^{-1}+\rep{1}_{1}^1+\rep{2}_{0}^0\Big]^a,
\end{align}
where the superscripts denote the R-symmetry $U(1)_L$ charges $C_L$.

\subsubsection{The Right $\mathcal{N}_R=0$ gauge theory}

The Right theory consists of one $\mathcal{N}=0$ vector potential $A_{\mu}^{A}$ and $q+2$ scalars $\phi^A$ in the adjoint of the Right  gauge group $G_R$ and $r$ Majorana spinors $\lambda^a$ and $2(q'+4)+t$ scalars $\Phi^a, \varphi^a$ in a  pseudoreal fundamental representation of  $G_R$.  The spacetime, global and gauge symmetries are given by
\be\label{Rsyms}
U(1)^{st}_R\times SO(q+2)\times F_q\times SO(q'+4)\times \mathcal{F}_{q'}\times SU(2)_R\times G_R,
\ee
  under which the various fields  transform as follows:
\begin{align}
A_{\mu}^{A}&:\Big[\rep{(1,1;1,1,1)}_{-2}+\rep{(1,1;1,1,1)}_{2}\Big]^A,\\
(q+2)\phi^A&:\Big[\rep{(q+2,1;1,1,1)}_0\Big]^A,\\
(r)\lambda^a&:\Big[(\rep{r;1,1,1})_{-1}+(\brep{r};\rep{1,1,1})_{1}\Big]^a,\\
2(q'+4)\Phi^a&:\Big[(\rep{1,1;q'+4,1,2})_0\Big]^a,\\
(t)\varphi^a&:\Big[(\rep{1,1;t,1})_0\Big]^a.
\end{align}
The groups $F_q$ and $\mathcal{F}_{q'}$ will be determined by the reality conditions. The fermion $\lambda$ representations  $\rep{r}=(\rep{spin}_{q+2},\rep{def}_{F_q})$  are given by the spinor  of $SO(q+2)$ and the  defining representation of $F_q$. The overall reality conditions imply that for $q\geq -1$ the flavour group is $F_q=S_q(P,\dot{P})$. For $q=-2$ the group $SO(q+2)$ disappears and the only possible flavour group consistent with overall reality is $F_{-2}=U(P)_F = SU(P)_F\times U(1)_F$, where the $U(1)_F$ charges carried by the fields will be denoted $C_F$. The scalars $\varphi$ are in the spinor of $SO(q'+4)$ and  defining of $\mathcal{F}_{q'}$, $\rep{t}=(\rep{spin}_{q'+4},\rep{def}_{F_q'})$. The overall reality conditions imply that for $q'\geq -3$ the flavour group is $\mathcal{F}_{q'}=S_{q'}(P',\dot{P'})$. For $q'=-4$ there are two options consistent with overall reality given by  $\mathcal{F}_{-4}=Sp(P)$ or $U(P)_\mathcal{F}=SU(P)_\mathcal{F}\times U(1)_\mathcal{F}$ , where the $U(1)_\mathcal{F}$ charges carried by the fields will be denoted $C_\mathcal{F}$.\\
\\
For the Right theory fields (see \autoref{indices} for a summary of the various spacetime, global and gauge indices),
\be
A_\mu^A, \qquad \phi^A_m, \qquad \lambda^{aI}_{\alpha x}, \qquad\Phi^{a\mathfrak{a}}_{M}, \qquad\varphi^{ai}_{X},
\ee
the most general Lagrangian consistent with the symmetries \eqref{Rsyms}  is given by,
\be\label{lagrang}
\begin{split}
\mathcal{L}=&-\frac{1}{4} F^{\mu\nu A}F_{\mu\nu}^A+\frac{1}{2} D_\mu\phi^A_m D^\mu\phi^A_m-\frac{1}{2} \lambda^{\alpha x}_{aI}(\gamma^\mu)_{\alpha}{}^{\beta} D_\mu\lambda^{aI}_{\beta x}+\frac{1}{2} D_\mu\Phi^{a\mathfrak{a}}_MD^\mu\Phi^{b\mathfrak{b}}_M\varepsilon_{\mathfrak{ab}}\Omega_{ab}+\frac{1}{2} D_\mu\varphi^{aX}_iD^\mu\varphi^{bi}_X\Omega_{ab}\\
&-\frac{g}{2}\lambda^{a\alpha x}_{I}(T^A)_{ab}\phi^A_m(\Gamma_m)_{x}{}^{y}(\gamma_{5})_{\alpha}{}^{\beta}\lambda^{bI}_{\beta y}-\frac{g^2}{4}\phi_m^A\phi_m^B\phi_n^C\phi_n^Df^{EAC}f^{EBD}\\
&+\Phi^{a\mathfrak{a}}_M\Phi^{b\mathfrak{b}}_M\Phi^{c\mathfrak{c}}_N\Phi^{d\mathfrak{d}}_NP_{\mathfrak{a}\mathfrak{b}\mathfrak{c}\mathfrak{d}abcd}+\varphi^{aX}_i\varphi^{biY}\varphi^{cZ}_{j}\varphi^{djW}\mathcal{P}_{XYZWabcd}+\Phi^{a\mathfrak{a}}_M\Phi^{b\mathfrak{b}}_N\varphi^{cX}_i\varphi^{di}_Y\varepsilon_{\mathfrak{a}\mathfrak{b}}\mathscr{P}_{XMNabcd}^Y,
\end{split}
\ee
where the relative coefficients of the fields generating the vector multiplet sector ($A, \phi, \lambda$) are fixed by regarding it as the dimensional reduction of a $D=q+6$ Yang-Mills theory coupled to $P+\dot{P}$ spinors.  The various invariant tensors are given explicitly by,
\begin{align}
P_{\mathfrak{a}\mathfrak{b}\mathfrak{c}\mathfrak{d}abcd}&=h_1T_{(\mathfrak{a}\mathfrak{b})(\mathfrak{c}\mathfrak{d})}M_{(ab)(cd)}+h_2T_{[\mathfrak{a}\mathfrak{b}][\mathfrak{c}\mathfrak{d}]}M_{[ab][cd]},\\
\mathcal{P}_{XYZWabcd}&=k_1\mathcal{T}_{(XY)(ZW)}M_{(ab)(cd)}+k_2\mathcal{T}_{[XY][ZW]}M_{[ab][cd]}+k_3\mathcal{T}_{(XY)(ZW)}M_{[ab][cd]}+k_4\mathcal{T}_{[XY][ZW]}M_{(ab)(cd)},\\
\mathscr{P}_{XMNabcd}^Y&=l_1\mathscr{T}_{(MN)X}^{Y}M_{[ab][cd]}+l_2\mathscr{T}_{[MN]X}^{Y}M_{(ab)(cd)},
\end{align}
where the free parameters $h_i, k_i, l_i$ are real and
\begin{align}
M_{(ab)(cd)}&=(T^A)_{ab}(T^A)_{cd}=\tau(-3f_{(abcd)}+\Omega_{a(c}\Omega_{d)b}),\\
M_{[ab][cd]}&=\Omega_{ab}\Omega_{cd}-2\Omega_{c[a}\Omega_{b]d},\\
T_{(\mathfrak{a}\mathfrak{b})(\mathfrak{c}\mathfrak{d})}&=\varepsilon_{\mathfrak{a}(\mathfrak{c}}\varepsilon_{\mathfrak{d})\mathfrak{b}},\\
T_{[\mathfrak{a}\mathfrak{b}][\mathfrak{c}\mathfrak{d}]}&=\varepsilon_{\mathfrak{a}\mathfrak{b}}\varepsilon_{\mathfrak{c}\mathfrak{d}}-2\varepsilon_{\mathfrak{c}[\mathfrak{a}}\varepsilon_{\mathfrak{b}]\mathfrak{d}},\\
\mathcal{T}_{(XY)(ZW)}&=S_{XY}S_{ZW}+2S_{Z(X}S_{Y)W}-2A_{Z(X}A_{Y)W},\\
\mathcal{T}_{[XY][ZW]}&=A_{XY}A_{ZW}-2A_{Z[X}A_{Y]W}+2S_{Z[X}S_{Y]W},\\
\mathscr{T}_{(MN)X}^{Y}&=\delta_{MN}\delta_{X}{}^{Y},\\
\mathscr{T}_{[MN]X}^{Y}&=(\Gamma_{MN})_{X}{}^{Y},\\
S_{XY}S_{ZW}&=\sum_{p}v_p(\Gamma_{M_1\dots M_p})_{(XY)}(\Gamma^{M_1\dots M_p})_{(ZW)},\\
A_{XY}A_{ZW}&=\sum_{p}u_p(\Gamma_{M_1\dots M_p})_{[XY]}(\Gamma^{M_1\dots M_p})_{[ZW]}.
\end{align}
Here, $\tau:=2\dim G_R/(\dim^2\rep{fund}_{G_R}+\dim\rep{fund}_{G_R})$. Note, $f_{(abcd)}$ is zero when the pseudoreal representation with index $a$ is the defining of $Sp(n)$. Examples of non-zero $f_{(abcd)}$ are given  by groups ``of type $E_7$'', for instance the $\rep{56}$ of $E_7$ \cite{Brown:1969, Borsten:2009zy, Marrani:2010de}.
 
\begin{table}[h!]
\begin{tabular}{C{2.5cm} C{2cm} C{2.5cm} C{2cm} C{2.5cm} C{2cm} C{2cm} C{2cm}}
\hline\hline
\\
~~Representation~~ & ~~$SO(1,3)_R^{st}$~~ & ~~$SO(q+2)$~~ & ~~$F_q$~~ & ~~$SO(q'
+4)$~~ & ~~$\mathcal{F}_{q'}$~~ & ~~$SU(2)_R$~~ & ~~$G_R$~~
\\ \\
\hline \\
~~Defining~~ & ~~$\mu,\nu,(+)$~~ & ~~$m,n,(-t_1^{q+2})$~~ & ~~$I,J,(s_q)$~~ & ~~$M,N,(+)$~~ & ~~$i,j,(t_0^{q'+4})$~~ & ~~$\mathfrak{a},\mathfrak{b}, (-)$~~ & ~~--~~
\\ \\
~~Fundamental~~ & ~~--~~ & ~~--~~ & ~~--~~ & ~~--~~ & ~~--~~ & ~~--~~ & ~~$a,b,(-)$~~
\\ \\
~~Adjoint~~ & ~~--~~ & ~~--~~ & ~~--~~ & ~~--~~ & ~~--~~ & ~~--~~ & ~~$A,B,(+)$~~
\\ \\
~~Spinor~~ & ~~$\alpha,\beta,(-)$~~ & ~~$x,y,(-t_0^{q+2})$~~ & ~~--~~ & ~~$X,Y,(-t_0^{q'+4})$~~ & ~~--~~ & ~~--~~ & ~~--~~
\\ \\
\hline
\hline
\end{tabular}
\caption{A summary of the representations/indices appearing in \eqref{lagrang}. Note, $(\text{sgn})$ indicates the sign picked up on raising/lowering a pair of contracted indices of that type.}\label{indices}
\end{table}

\subsection{The $[\mathcal{N}_L=2] \times [\mathcal{N}_R=0]$ product}\label{prodsec}

The product of these two multiplets yields $\mathcal{N}=2$ supergravity coupled to $(1+q+2+r)$ vector multiplets and $(q'+4+t/2)$ hyper multiplets, whose origin can be traced through,
\be\label{genprod}
\begin{split}
\textit{Left}\otimes\textit{Right}&=\Big[\rep{V}_2^A\oplus\rep{C}_2^a\Big]\otimes\Big[V^A\oplus(q+2)\phi^A\oplus(r)\lambda^a\oplus2(q'+4)\Phi^a\oplus(t)\varphi^a\Big]\\
&=\rep{V}_2^A\otimes\Big[V^A\oplus(q+2)\phi^A\Big]\oplus\rep{C}_2^a\otimes \Big[(r)\lambda^a\oplus2(q'+4)\Phi^a\oplus(t)\varphi^a\Big]\\
&=\rep{G}_2\oplus(1+q+2+r)\rep{V}_2\oplus(q'+4+t/2)\rep{H}_2.
\end{split}
\ee
The supergravity theory  inherits the  global symmetries
\be\label{totsys}
U(1)^{st}\times H \times \mathcal{H}=U(1)^{st}\times SO(q+2)\times F_q\times SO(q'+4)\times \mathcal{F}_{q'}\times SU(2)_L\times SU(2)_R\times U(1)_L\times U(1)_-\ee
 directly from the two Yang-Mills factors. Noting that the vector multiplets carry non-trivial $SO(q+2)\times F_q$ representations, it is clear that  the corresponding SK manifold $G/H$ will have  $SO(q+2)\times F_q\subset H$. Similarly,   the hyper multiplets carry non-trivial $SO(q'+4)\times \mathcal{F}_{q'}\times SU(2)_R$ representations and therefore $SO(q'+4)\times \mathcal{F}_{q'}\times SU(2)_L \times SU(2)_R\subset \mathcal{H}$ will contribute to the Q manifold $\mathcal{G}/\mathcal{H}$.\\
 \\
Some more detailed comments on the various $U(1)$ factors appearing in \eqref{totsys} are in order. First, the $U(1)^{st}$ and  $U(1)_-$ charges, denoted $C^{st}, C_-$ respectively,   are given by the sum and difference of the Left and Right helicities $C^{st}_L, C^{st}_R$:
\begin{align}
C^{st}&=C^{st}_L+C^{st}_R,\\
C_-&=C^{st}_L-C^{st}_R.
\end{align}
Unlike the symmetries inherited directly and independently from the gauge factors, the ``helicity difference'' $U(1)_-$ is not a priori a symmetry of the gravitational theory.  However, as noted in \cite{ Anastasiou:2013hba} if the scalar manifold of the supergravity theory is symmetric the $U(1)_-$ symmetry is required by the squaring procedure. The simplest example is given by the product of two $\mathcal{N}=0$ gauge potentials yielding axion-dilaton gravity, where the axion and dilaton parametrise $SL(2, \R)/U(1)_-$. Similarly, for maximal supersymmetry we have a global $SU(8)$ in  $\mathcal{N}=8$ supergravity, which factors into $U(1)_-\times SU(4)_L\times SU(4)_R$, where $SU(4)_{L/R}$ are the R-symmetries of the Left and Right $\mathcal{N}=4$ Yang-Mills theories. On the other hand, if the scalar manifold is non-symmetric the $U(1)_-$ \emph{cannot} be present. Note, this is reflected precisely by the double-copy construction\footnote{We thank Henrik Johansson for illuminating discussions regarding this point.}: if and only if  the scalar manifold is symmetric are the gravity amplitudes  $U(1)_-$ invariant, as made evident by the examples constructed in \cite{Chiodaroli:2015wal}.\\
\\
To make this distinction clear in the present context  we briefly illustrate  the appearance of  $U(1)_-$  in the symmetric generic Jordan sequence, where it  is identified with the axio-dilatonic $%
U(1)_{g}$ in the stabilizer of \eqref{jordan}. Note, using the magic Jordan algebraic embedding $%
\mathfrak{J}_{3}^{\mathbb{A}}\supset \mathbb{R}\oplus \mathfrak{J}_{2}^{%
\mathbb{A}}$, considered at the level of conformal symmetries, the second enhancement can be simultaneously   made manifest at the expense of restricting to $q=1,2,4,8$. Adopting this starting point we then further branch the axio-dilatonic $SU(1,1)\cong SL(2, \mathbb{R})$ to its
non-compact Cartan, which generates the 5-grading:
\begin{eqnarray}
Conf\left( \mathfrak{J}_{3}^{\mathbb{A}}\right)  &\supset &Conf\left(
\mathbb{R}\oplus \mathfrak{J}_{2}^{\mathbb{A}}\right) \times S_{q}\simeq\label{magenhance}
SU(1,1)\times SO(q+2,2)\times S_{q}  \nonumber \\
&\supset& SO(1,1)\times SO(q+2,2)\times S_{q};\label{br-1} \\
&&  \nonumber \\
\mathbf{Adj} &=&\left(
\mathbf{1}, \mathbf{Adj}, \mathbf{1}\right) +(\mathbf{3},\mathbf{%
1,1})+\left( \mathbf{1},\mathbf{1},\mathbf{Adj}\right) +(\mathbf{2},%
\mathbf{spin}, \mathbf{def})  \nonumber \\
&=&\left( \mathbf{Adj}, \mathbf{1}\right) ^{0}+\left( \mathbf{1},%
\mathbf{1}\right) ^{0}+\left( \mathbf{1}, \mathbf{Adj}%
\right) ^{0}+(\mathbf{1},\mathbf{1})^{-2}+(\mathbf{spin}\mathbf{,def}%
)^{-1}+(\mathbf{spin}\mathbf{,def})^{+1}+(\mathbf{1},%
\mathbf{1})^{+2}.\label{br-2}
\end{eqnarray}%
One  recognises  the $G$ of  homogeneous non-symmetric  
projective SK manifolds, given in  \eqref{SKnon}, as  the \textit{non-negatively} graded part of (\ref{br-2}), with  the global $SU(2)$ factor omitted. However, for the symmetric case we also have the additional negative grade $(\mathbf{1},\mathbf{1})^{-2}$ component, which when linearly composed with  the   $(\mathbf{1},\mathbf{1}%
)^{+2}$ component  yields a maximal compact $U(1)$ subgroup of  $SU(1,1)$, identified with $U(1)_-$, generating the enhancement  $SO(1,1)\times SO(q+2,2) \longrightarrow SU(1,1)\times SO(q+2,2)$. For the magic cases of $q=1,2,4,8$  we see that \eqref{magenhance} and \eqref{br-2} imply the further enhancement $SU(1,1)\times SO(q+2,2)\times
S_{q} \longrightarrow Conf\left( \mathfrak{J}_{3}^{\mathbb{A}}\right)$. In this sense, \textit{at least} within the cubic models, the fact that the
extra $U(1)$ is missing in the $T^{3}$ model can be traced back to the fact
that the $T^{3}$ model does not contain (as a truncation) any element of the
generic Jordan sequence.

\subsubsection{Vector multiplets}\label{vectors}

In order to reproduce the SK manifolds of \autoref{vector}, we consider those factors of \eqref{genprod} contributing to the vector multiplet sector, specifically:
\be\label{vectorsgen}
\rep{V}_2^A\otimes\Big[V^A\oplus(q+2)\phi^A\Big]\oplus\rep{C}_2^a\otimes (r)\lambda^a=\rep{G}_2\oplus(1+q+2+r)\rep{V}_2.
\ee
The resulting content $\rep{G}_2\oplus(1+q+2+r)\rep{V}_2$  carries non-trivial representations, inherited directly from the Left and Right Yang-Mills multiplets, under the 
\be
U(1)^{st}\times SO(q+2)\times F_q\times SU(2)_L\times U(1)_L\times U(1)_-
\ee
 factor of \eqref{totsys},
\begin{align}\label{g2vect}
&\rep{(1,1,1)}_{-4}^{(0,0)}+\rep{(1,1,1)}_{4}^{(0,0)}\\
\nonumber\\
&\rep{(1,1,2)}_{-3}^{(1,1)}+\rep{(1,1,2)}_{3}^{(-1,-1)}\nonumber\\
\nonumber\\
&\rep{(1,1,1)}_{-2}^{(2,2)}+\rep{(1,1,1)}_{2}^{(-2,-2)}&&+&&\rep{(1,1,1)}_{-2}^{(-2,2)}+\rep{(1,1,1)}_{2}^{(2,-2)}&&+&&\rep{(q+2,1,1)}_{-2}^{(0,-2)}+\rep{(q+2,1,1)}_{2}^{(0,2)}\nonumber\\
&&&+&&(\rep{r,1})_{-2}^{(-1,0)}+(\brep{r},\rep{1})_{2}^{(1,0)}\nonumber\\
\nonumber\\
&&&&&\rep{(1,1,2)}_{-1}^{(-1,3)}+\rep{(1,1,2)}_{1}^{(1,-3)}&&+&&\rep{(q+2,1,2)}_{-1}^{(1,-1)}+\rep{(q+2,1,2)}_{1}^{(-1,1)}\nonumber\\
&&&+&&(\rep{r,2})_{-1}^{(0,1)}+(\brep{r},\rep{2})_{1}^{(0,-1)}\nonumber\\
\nonumber\\
&&&&&\rep{(1,1,1)}_{0}^{(0,4)}+\rep{(1,1,1)}_{0}^{(0,-4)}&&+&&\rep{(q+2,1,1)}_{0}^{(2,0)}+\rep{(q+2,1,1)}_{0}^{(-2,0)}\nonumber\\
&&&+&&(\rep{r,1})_{0}^{(1,2)}+(\brep{r},\rep{1})_{0}^{(-1,-2)}.\nonumber
\end{align}
Leaving the $T^3$ model aside for the moment,  at this stage we are able to reproduce all homogeneous SK manifolds of \autoref{vectors} by adjusting the field multiplicities, symmetries and representations (and hence couplings) of the Right Yang-Mills theory.   The choices of $q$ and $F_q$ (and the corresponding representation $\rep{r}$) giving the non-symmetric (\autoref{nonsym}), generic Jordan (\autoref{genJ}), magic (\autoref{vectormagic}) and minimally coupled (\autoref{mcoup}) theories are summarised  in \autoref{vectortab}. Note, the first three choices  reproduce the previous BCJ double-copy construction of the   non-symmetric, generic Jordan and magic theories appearing in \cite{Chiodaroli:2015wal}. Note, as originally observed in \cite{Chiodaroli:2015wal} the minimally coupled sequence can also be obtained as a truncation of the generic Jordan sequence and is thus already included, in this sense, in the double-copy construction of the generic Jordan sequence.\\
\\
The semi-simple symmetries and their representations are matched directly to those of the corresponding supergravity theories. The only minor subtlety is the correct identification of the various $U(1)$ charges.  Including the additional $U(1)_-$, the correct charges for each $U(1)$ factor appearing in the gravitational theory are  given by an invertible linear combination of  $C_-$ with $C_L$ and $C_F$, which are inherited directly from the  Left and Right factors respectively, as described in the final column of   \autoref{vectortab}. In these cases, all scalar fields appearing in the supergravity theory transform under a manifest global symmetry, which is sufficient to ensure that the scalar manifold is (locally) homogeneous \cite{ArkaniHamed:2008gz, Chiodaroli:2015wal}.  By contrast, comparing the $\rep{(1,1,1)}^{x}_{0}+\rep{(1,1,1)}^{-x}_{0}$ terms of  \eqref{nonsymvect}  and \eqref{g2vect} we see that  $U(1)_-$ is \emph{not} a symmetry of the non-symmetric  theory and must be discarded. This accords perfectly with the double-copy construction of \cite{Chiodaroli:2015wal}. In the absence of the  $U(1)_-$ the  vanishing of the single soft-dilaton/axion limits must be checked independently to establish (local) homogeneity.\\
\begin{table}[h!]
\begin{tabular}{C{2.5cm} C{2cm} C{2cm} C{3.5cm} L{7.75cm} }
\hline\hline
\\
~~$(q,P,\dot{P})$~~ & ~~$F_q$~~ & ~~$\rep{r}$~~ & ~~Theory~~ & ~~Comments~~
\\ \\
\hline \\
~~$(\geq -1,P,\dot{P})$~~ & ~~$S_q(P,\dot{P})$~~ & ~~\autoref{SqPPreps}~~ & ~~Non-symmetric~~ & ~~$SU(2)=SU(2)_L$, $C=C_L$, drop $U(1)_-$~~
\\ \\
~~$(\geq -1,0,0)$~~ & ~~-//-~~ & ~~-//-~~ & ~~Generic Jordan~~ & ~~$SU(2)=SU(2)_L$, $C=C_L$, $C_g=C_-$~~
\\ \\
~~$(n,1,0)$~~ & ~~$S_n(1,0)$~~ & ~~\autoref{Jreps}~~ & ~~Magic~~ & ~~$SU(2)=SU(2)_L$, $C'=2C_L+C_-$, $C''=-C_L+C_-$~~
\\
&&&& ~~Enhancement to $\rep{d}$ after squaring~~
\\ \\
~~$(-2,P,0)$~~ & ~~$U(P)_F$~~ & ~~$\rep{P}^{-1}+\brep{P}^{1}$~~ & ~~Minimally coupled~~ & ~~$SU(2)=SU(2)_L$, $C=C_L+C_F$~~
\\
&&&& ~~$C_{min}=(P/2)C_L+(1+P/2)C_--(2+P/2)C_F$~~
\\
&&&& ~~$C_P=(P/4)C_L-(P/4)C_--(1+P/4)C_F$~~
\\ \\
~~-//-~~ & ~~-//-~~ & ~~-//-~~ & ~~$T^3$~~ & ~~$SU(2)=SU(2)_L$, $C_T=2C_L-C_-$~~
\\
&&&& ~~No. of $U(1)$'s not conserved~~
\\ \\
\hline
\hline
\end{tabular}
\caption{The choice of $(q,P,\dot{P})$ and $F_q$ for the Right Yang-Mills factor and the required linear combinations of $U(1)$ charges leading to the non-symmetric (\autoref{nonsym}), generic Jordan (\autoref{genJ}), magic (\autoref{vectormagic}) and minimally coupled (\autoref{mcoup}) $\N=2$ supergravity theories coupled to vector multiplets.}\label{vectortab}
\end{table}\\
Finally, let us briefly comment on the absence of the $T^3$ model. Note, the $P=0$ minimally coupled model has the same field content and  scalar coset as the $T^3$ model, but with different representations as can be seen by  comparing \eqref{mcreps} with \eqref{T3reps}.  In particular, the two vector potentials and their duals transform as the $\rep{4}$ and $\rep{2}+\rep{2}$ of $SL(2, \R)$ for the $T^3$ and minimally coupled model, respectively. Using the same starting point as the minimally coupled model, but letting $C_T=2C_L-C_-$ while dropping entirely the second independent $U(1)$, one can reorganise the content into that of the $T^3$ model. Recall, however, that since the $T^3$ model has a symmetric manifold the global $U(1)_-$ is present and we are therefore actually throwing away a symmetry that is inherited from the gauge factors.  In terms of the double-copy there will be non-trivial amplitudes in the $T^3$ model that are \emph{not} generated by the Yang-Mills factors, essentially because we are relaxing the second $U(1)$ symmetry present in the minimally coupled starting point.   To give another perspective, the squaring and  double-copy constructions  are many-to-one; there is no way to pass from minimally coupled to $T^3$. This leaves the possibility of using two $\mathcal{N}=1$ Yang-Mills multiplets, but from \autoref{prods} we see immediately that this generates  at least one hyper multiplet. The apparent absence of only the $T^3$ model amongst all homogeneous $\N=2$ supergravities coupled to vector multiplets is rather surprising.  Although, it should be recalled that the $T^3$ model stands alone in the sense that its five dimensional origin is \emph{pure} minimal supergravity, which itself does not admit a squaring origin. Moreover it is an isolated case in  the classification of symmetric projective special K\"ahler manifolds \cite{deWit:1991nm}. The final logical possibility is that the product of two Yang-Mills theories can have more supersymmetry than the sum of its factors. We leave such speculations for future consideration.

\subsubsection{Hyper multiplets}\label{hypers}

In order to reproduce the Q manifolds of \autoref{hyper} let us isolate   those terms generating  hyper multiplets in \eqref{genprod},
\be\label{hepergen}
\rep{C}_2^a\otimes\Big(2(q'+4)\Phi^a\oplus(t)\varphi^a\Big)=(q'+4+t/2)\rep{H}_2,
\ee
 and label the resulting content under $U(1)^{st}\times SO(q'+4)\times \mathcal{F}_{q'}\times SU(2)_R\times SU(2)_L$. Of course the fields will also be charged under $U(1)_L\times U(1)_-$, but since these factors  are absorbed by  the SK manifolds of the vector multiplet sector we omit them here.\\
\\
 The resulting content carries the representations:
\begin{align}\label{genhyprep}
&\rep{(q'+4,1,2,1)}_{-1}+\rep{(q'+4,1,2,1)}_{1}&&+&&\rep{(t,1,1)}_{-1}+\rep{(t,1,1)}_{1}\\
&\rep{(q'+4,1,2,2)}_{0}&&+&&\rep{(t,1,2)}_{0}.\nonumber
\end{align}
 As for the vector sector   we are able to reproduce all homogeneous Q manifolds of \autoref{hyper} by adjusting the field multiplicities, symmetries and representations of the Right Yang-Mills theory.   The choice of $(q',P',\dot{P}')$ and  $\mathcal{F}_{q'}$ for the Right Yang-Mills factor  (and the corresponding representation $\rep{t}$) giving the non-symmetric (\autoref{hnonsym}), generic Jordan (\autoref{hgenJ}), magic (\autoref{hvectormagic}), minimally coupled (\autoref{hmcoup}), projective quaternionic (\autoref{hproj}) and exceptional $T^3$ (\autoref{hT3}) theories are summarised  in  \autoref{hypertab}.\\
\\

\begin{table}[h!]
\begin{tabular}{C{2.5cm} C{2cm} C{2cm} C{3.5cm} L{8cm} }
\hline\hline
\\
~~$(q',P',\dot{P'})$~~ & ~~$\mathcal{F}_{q'}$~~ & ~~$\rep{t}$~~ & ~~Theory~~ & ~~Comments~~
\\ \\
\hline \\
~~$(\geq -3,P',\dot{P'})$~~ & ~~$S_{q'}(P',\dot{P'})$~~ & ~~\autoref{Hyprep}~~ & ~~Non-symmetric~~ & ~~$SU(2)_{ns}=SU(2)_L\times SU(2)_R$~~
\\ &&&& ~~The diagonal identification causes the breaking~~
\\ &&&& ~~$SO(q'+4)\rightarrow SO(q'+3)$ and thus $\rep{t}\rightarrow \rep{s}$~~
\\ \\
~~$(\geq -3,0,0)$~~ & ~~-//-~~ & ~~-//-~~ & ~~Generic Jordan~~ & ~~$SU(2)=SU(2)_L$, $SU(2)'=SU(2)_R$~~
\\ \\
~~$(n,1,0)$~~ & ~~$S_n(1,0)$~~ & ~~\autoref{maghyptab}~~ & ~~Magic~~ & ~~$SU(2)=SU(2)_L$, $SU(2)'=SU(2)_R$~~
\\ &&&& ~~Enhancement to $\rep{f}$ before squaring~~
\\ \\
~~$(-4,P,0)$~~ & ~~$U(P)$~~ & ~~$\rep{P}^{-1}+\brep{P}^{1}$~~ & ~~Minimally coupled~~ & ~~$SU(2)=SU(2)_L$, $SU(2)_R$ drops out automatically~~
\\ \\
~~$(-4,P,0)$~~ & ~~$Sp(P)$~~ & ~~$\rep{2P}$~~ & ~~Projective quaternionic~~ & ~~$SU(2)=SU(2)_L$, $SU(2)_R$ drops out automatically~~
\\ \\
~~-//-~~ & ~~$SU(2)_E$~~ & ~~$\rep{4}$~~ & ~~Exceptional~~ & ~~$SU(2)=SU(2)_L$~~
\\ \\
\hline
\hline
\end{tabular}
\caption{The choice of $(q',P',\dot{P}')$ and $\mathcal{F}_{q'}$ for the Right Yang-Mills factor and the required identifications $SU(2)_{L/R}$  leading to the non-symmetric (\autoref{hnonsym}), generic Jordan (\autoref{hgenJ}), magic (\autoref{hvectormagic}), minimally coupled (\autoref{hmcoup}), projective quaternionic (\autoref{hproj}) and exceptional $T^3$ (\autoref{hT3}) theories.}\label{hypertab}
\end{table}
\noindent In this case all those theories with symmetric quaternionic manifolds follow straightforwardly from the Right factor since the full isometry group is given by $\mathcal{H}=SU(2)_L\times SO(q'+4)\times \mathcal{F}_{q'}\times SU(2)_R$ or its relevant enhancement, which occurs already in the Right factor \emph{before} squaring. See for example the magic cases given in \autoref{maghyptab}. For instance, for the octonionic magic theory the Right factor scalars $\Phi^a$ and $\varphi^a$ transform irreducibly under  the enhanced $E_{7}\cong [Conf(\mathfrak{J}^{\Al_\C}_{3})]_c$ symmetry of the Right factor, 
\be
\begin{split}
SO(12) \times SU(2)'&\hookrightarrow E_{7};\\
(\rep{12, 2})+(\rep{32, 1})&\longrightarrow\rep{56},
\end{split}
\ee
where $SO(q'+4)\times \mathcal{F}_{q'}|_{q'=8}\cong SO(12)$ and   $SU(2)'\cong SU(2)_R$.  From \eqref{magicbreak1} we see that in total the hyper scalars transform as 
\be
(\rep{56, 1})^1+(\rep{56, 2})^0+(\rep{56, 1})^{-1}
\ee
under the combined Left/Right symmetries 
$\mathcal{H}\cong[Conf(\mathfrak{J}^{\Oct_\C}_{3})]_c\times SU(2)$ where $SU(2)\cong SU(2)_L$. The remaining cases are summarised in \autoref{hypertab}, with the appropriate identification of the $SU(2)$ factors.\\ 
\\
The non-symmetric family is a little more subtle.  On the gravitational side  the hyper multiplet scalars that are singlets under the $SO(q+3)\times S_q(P,\dot{P})$ subgroup of $\mathcal{H}=SO(q+3)\times S_q(P,\dot{P}) \times SU(2)_{ns}$ transform under  $\mathcal{H}$ as 
\be\label{desired}
(\rep{1,1,3})+(\rep{1,1,1}),
\ee
as can be seen from \eqref{hypref}. On the other hand, from   \eqref{genhyprep} we observe that the only 
\be
SO(q+3)\times S_q(P,\dot{P})\subset SO(q'+4)\times S_q(P,\dot{P})
\ee
 singlets in hyper multiplet scalar sector that follow from  the product of the Left/Right factors \eqref{hepergen} transform  as a
$
 \rep{(1,1, 2, 2)}
 $
under the Left/Right symmetries $SO(q+3)\times S_q(P,\dot{P})\times SU(2)_R\times SU(2)_L$. Note, the $\rep{2}$ of $SU(2)_L$ is required by R-symmetry and hence the only way \eqref{desired} may be reproduced is by identifying $SU(2)_{ns}$ with a diagonal subgroup of  $SU(2)_R\times SU(2)_L$,
\be
SU(2)_{ns} \cong [SU(2)_R\times SU(2)_L]_{\text{diag}} \supset SU(2)_R\times SU(2)_L,
\ee
under which
\be
 \rep{(1,1, 2, 2)} \rightarrow  \rep{(1,1, 2\times2)}= (\rep{1,1,3})+(\rep{1,1,1}).
\ee
The observation that the symmetric Q manifolds have an extra $SU(2)$ with respect to the non-symmetric case follows from an argument analogous to the treatment of the extra $U(1)_-$ appearing  in the symmetric SK manifolds with respect to the non-symmetric  SK manifolds.
In order to see this, we start once again from the Jordan algebraic
embedding $\mathfrak{J}_{3}^{\mathbb{A}}\supset \mathbb{R}\oplus \mathfrak{J}%
_{2}^{\mathbb{A}}$, considered now at the level of quasi-conformal symmetries,
and then further branch it in order to obtain an $SO(1,1)$ generating the
5-grading (here $S_{q}\simeq S_{q}(1,0)\simeq S_{q}(0,1)$, where $q=1,2,4,8$):%
\begin{eqnarray}\label{maghypenhance}
QConf\left( \mathfrak{J}_{3}^{\mathbb{A}}\right)  &\supset &QConf\left(
\mathbb{R}\oplus \mathfrak{J}_{2}^{\mathbb{A}}\right) \times S_{q}\simeq
SO(q+4,4)\times S_{q}  \nonumber \\
&\supset &SO(1,1)\times SO(q+3,3)\times S_{q};\label{br-3} \\
&&  \nonumber \\
\mathbf{Adj} &=&\left(
\mathbf{Adj},\mathbf{1}\right) +\left( \mathbf{1},\mathbf{Adj}%
\right) +(\mathbf{spin}\mathbf{,def})  \nonumber \\
&=&\left( \mathbf{Adj},\mathbf{1}\right) ^{0}+(\mathbf{1},%
\mathbf{1})^{0}+\left( \mathbf{1},\mathbf{1},\mathbf{Adj}\right) ^{0}
\nonumber \\
&&+\left( \mathbf{q+6},\mathbf{1}\right) ^{-2}+(\mathbf{spin}^{\prime }\mathbf{,def})^{-1}+(\mathbf{spin}\mathbf{%
,def})^{+1}+\left( \mathbf{q+6},\mathbf{1}\right) ^{+2}.\label{br-4}
\end{eqnarray}%
One recognizes that $\mathcal{G}$ of the homogeneous non-symmetric class of
quaternionic manifolds is given by the \textit{non-negatively} graded part
of the branching (\ref{br-4}). The aforementioned extra $%
SU(2)$  requires in addition the negatively graded part of (\ref{br-4}). The 
two negative grade terms can be present only in the symmetric case, where they generate the   enhancement  from $SO(1,1)\times SO(q+3,3)$  to $SO(q+4,4)$. From \eqref{maghypenhance} we observe that in the magic cases a second enhancement takes place,
 $SO(q+4,4)\times S_{q}\longrightarrow $ $QConf\left(
\mathfrak{J}_{3}^{\mathbb{A}}\right) $ (recall, we have restricted to the special values of $q=1,2,4,8$ here to make this manifest). Note that the two $SU(2)$ 
factors are generated by the $so(4)$  in the maximal compact subalgebra  $so(q+4)\oplus so(4)\subset so(q+4,4)$, while the unique $SU(2)_{ns}$ appearing in the   non-symmetric homogeneous
manifolds is generated by the $so(3)$ summand in $so(q+3)\oplus so(3)\subset  so(q+3,3)$. Although it is not a priori obvious why the restriction $SU(2)_{ns}\subset SU(2)_R\times SU(2)_L$ would be required by the double-copy  construction, the above symmetry arguments strongly suggest it will be effected at the level of amplitudes.

\section{Conclusion}

We have shown that all ungauged $D=4$, $\N=2$ supergravity theories with homogeneous scalar manifolds are the square of Yang-Mills with two isolated exceptions, pure $\N=2$ supergravity and the $T^3$ model in the vector sector. This completes the classification of all supergravities with eight or more supercharges and homogeneous scalar manifolds in $D\geq 4$  with a Yang-Mills squared origin  (up to the possibility that more subtle, yet to be appreciated, mechanisms may enter the game).  There are two obvious directions for future work  (i) $D=4$, $\N=2$ supergravities with non-homogeneous scalar manifolds and (ii) $D=4$, $\N=1$ supergravities, first with homogeneous (non)symmetric and then non-homogeneous scalar manifolds. We should emphasise, however, that at present it is unclear how one would proceed in the non-homogeneous cases. Finally, one can consider the BCJ colour-kinematic duality compatibility of the Right Yang-Mills factor to determine whether or not all examples appearing in the classification are indeed double-copy constructible. For the non-symmetric hyper multiplet sector we would anticipate a BCJ origin for the identification of the Left and Right $SU(2)$ factors, which at present is not understood.  

\acknowledgements

We are grateful to Marco Chiodaroli and Henrik Johansson  for useful discussions. We would like to thank Gerard `t Hooft and Antonino Zichichi, directors of the 55th International School Of Subnuclear Physics,  EMFCSC, Erice, Sicily, for the stimulating atmosphere in which the final stages of the project were conceived. The work of LB is supported by a Schr\"odinger Fellowship. SN is supported by a Leverhulme Research Project Grant.  MJD is grateful to the Leverhulme Trust for an Emeritus Fellowship and to Philip Candelas for hospitality at the Mathematical Institute, Oxford.  This work was  supported by the STFC under rolling grant ST/G000743/1.

\appendix

\section{Enhancements}\label{enhancements}

In this appendix we address the problem of enhancements. It should be obvious from equation \eqref{totsys} that upon squaring, we take the direct product between the global internal symmetry groups of the two Yang-Mills sides which can schematically be expressed as $Sym_L\times Sym_R\times U(1)_-$ in the case of a symmetric scalar coset. Then \autoref{vectortab} and \autoref{hypertab} give the explicit rules of how to go from this direct product of groups to the $H\times \mathcal{H}$ of the desired supergravity theory. In the non-symmetric cases one needs to drop $U(1)_-$ and identify  the two $SU(2)_{L/R}$. In the symmetric cases things differ between the hyper and vector multiplet sectors. In the hyper multiplet scalar sector the full isometry group is given by $\mathcal{H}=SU(2)_L\times Sym_R$. However, in the vector multiplet scalar sector there are cases where the isometry group $H$ is given directly by the remaining factors of $Sym_L\times Sym_R\times U(1)_-$  and   other cases where these need to be further enhanced in order to form the full $H$. The purpose of this section is to understand when this enhancement occurs in the symmetric vector multiplet sector and study its Yang-Mills origin.\\
\\
To answer the question of \emph{when} such enhancements occur in the symmetric cases it is enough to observe that (1) they never occur in the hyper sector and (2) they do not occur in the generic-Jordan vector sector series. Having made this observation the answer can be summarised as follows:\\
\\
\textit{Whenever the scalars parametrising a symmetric coset space originate from both ``boson $\otimes$ boson'' and ``fermion $\otimes$ fermion'' terms,  an enhancement is required.}\\
\\
In particular, the hyper sector  scalars  always have  a purely ``\textit{boson $\otimes$ boson}'' origin and no enhancement is required, consistent with the identification $\mathcal{H}=SU(2)_L\times Sym_R$. Similarly, in the generic Jordan vector sector series $P=\dot{P}=0$ so that the Right theory has no fermions, again implying that no enhancement is required. \\
\\
The more interesting question is how, when needed,  these enhancements arise in terms of  the Yang-Mills factors:\\
\\
\textit{What is the Yang-Mills origin of the extra generators required to enhance the symmetries to the full $H$ isometry group?}\\
\\
This question was addressed in the context of squaring pure super-Yang-Mills theories in \cite{Borsten:2013bp, Anastasiou:2013hba, Anastasiou:2015vba}. It is instructive to recall the problem through the paradigmatic example of $\mathcal{N}=8$ supergravity as the product of two pure $\mathcal{N}=4$ super-Yang-Mills, each having an $SU(4)$ global internal R-symmetry. The problem there was to find the missing generators required to enhance $SU(4)\times SU(4)\times U(1)_-\subset SU(8)$. The adjoint decomposition goes as:
\be
\rep{63}\rightarrow\rep{(15,1)}^0+\rep{(1,15)}^0+\rep{(1,1)}^0+(\rep{4},\brep{4})^1+(\brep{4},\rep{4})^{-1},
\ee
which means that the missing generators should carry defining and conjugate-defining indices with respect to the R-symmetry groups. Furthermore the generators should act simultaneously on both factors and should turn bosonic into fermionic states and vice versa. All these features led us to propose that the missing generators where composed by the tensor product of the Left and Right supersymmetry generators. Indeed, the tensor product of the Left and Right supercharges provided a precise guide to constructing the generators. The caveat here is that strictly speaking the generators cannot be those of supersymmetry because the momentum factors arising from the partial derivatives should be removed by hand. This caveat can be equivalently thought as the problem of trying to build a dimensionless bosonic parameter from the product of two supersymmetry parameters.\\
\\
As a concrete example of the theories studied in this paper we can focus on the octonionic magic vector multiplet sector. We need to find the Yang-Mills origin of the enhancement $SO(10)\times U(1)\subset E_6$ where the adjoint representation decomposes as:
\be
\rep{78}\rightarrow\rep{45}^0+\rep{1}^0+\rep{16}^3+\brep{16}^{-3}.
\ee 
It is instructive to notice that while the simultaneous supersymmetry picture now fails, since the Right theory is non-supersymmetric, the missing generators still carry representations similar to those of the spinors and still need to mix bosonic with fermionic sates. It should be emphasised that although  the combined Left/Right generators correspond to a legitimate bosonic symmetry transformation on the supergravity side, the individual generators themselves to not induce a symmetry transformation on the Yang-Mills factors individually.\\
\\
In the Left sector, the vector multiplet is described by the on-shell superfield,
\be 
V_{2-}=\bar{\phi}+\eta^{\mathfrak{a}}\psi_{\mathfrak{a}-}+\eta^1\eta^2 V_-,
\ee  
and similarly for $V_+$. For the half-hyper, we have,
\be
C_2=\chi_+ +\eta^{\mathfrak{a}}\sigma_{\mathfrak{a}} +\eta^1\eta^2\chi_-,
\ee
where the gauge indices have been omitted for simplicity. The action of the desired ladder operator on the superfields is simply:
\be
\begin{split}
\left[L_L\right]_- V_{2+}&= C_2,  \\
\left[L_L\right]_-C_2&=V_{2-},\\
\left[L_L\right]_- V_{2-}&=0,
\end{split} 
\ee
and similarly for $\left[L_L\right]_+$. It should be noted that the operator carries both adjoint and defining gauge indices which are contracted accordingly with the superfield it acts on. Using the standard (anti)-commutation relations, we can write the ladder operator  as
\be
 \left[L_L\right]_-:=\int\widetilde{dp}\sum_{k=0}^2\left[
 C_{2\{k\}} (V_{2+\{k\}})^\dagger
+ V_{2-\{k\}} (C_{2\{k\}})^\dagger
\right],
\ee
where we defined $S_{\{k\}}\equiv\frac{\partial}{\partial^k\eta}S|_{(\eta=0)}$.\\
\\
The states of the Right sector can be written as
\be 
\label{statesN0}
 A_+,\quad\lambda_{+x}^I, \quad \phi^m, \quad\lambda_{-I}^{x},\quad A_-,
\ee
where again the gauge indices have been omitted. The action of the desired ladder operator on the states is:
\be
\begin{aligned}
\left[L_R\right]^I_{-x} A_+
&=\lambda^I_{+x},\\
\left[L_R\right]^I_{-x}\lambda^J_{+y}
&=\phi^m (\gamma_m)_{xy} \mathcal{I}^{IJ},\\
\left[L_R\right]^I_{-x} \phi^m
&=\lambda_{-J}^y(\gamma^m)_{yx}\mathcal{I}^{IJ},\\
\left[L_R\right]^I_{-x}\lambda_{-J}^{y}
&= A_-\delta_x^y \delta^{I}_{J},\\
\left[L_R\right]^I_{-x} A_-&=0,
\end{aligned}
\ee
and similarly for the raising operator $\left[L_R\right]_{+I}^x$. In the minimally coupled cases $\phi^m=0$. In the magic cases $\mathcal{I}^{IJ}=\id$ for $n=1,2,8$ and $\mathcal{I}^{IJ}=\varepsilon^{IJ}$ for $n=4$. We can use the standard (anti)-commutation relations,
\be
\begin{aligned}
\left[ A_+(p),A^\dagger_+(q)\right]&=(2\pi)^32E_p\delta^3(\vec{p}-\vec{q}),\\
\{\lambda_{+x}^{I}(p),(\lambda^\dagger_+)_{J}^{y}(q)\}&=(2\pi)^32E_p\delta^3(\vec{p}-\vec{q})~\delta^y_x \delta^{I}_J,\\
\left[ \phi^m(p), \phi^{n\dagger}(q)\right]&=(2\pi)^32E_p\delta^3(\vec{p}-\vec{q})~\delta^{mn},
\end{aligned}
\ee
to pack the ladder operator in the simple form
\be
\left[L_R\right]^{I}_{-x}:=\int\widetilde{dp}\left[-\lambda_{+x}^{I}(A_{+})^\dagger
+\phi^m (\gamma_m)_{xy} (\lambda^{I}_{+y})^\dagger
+\lambda_{-}^{Iy}(\gamma_m)_{xy}(\phi^m)^\dagger
 -A_{-}(\lambda_{-}^{Ix})^\dagger\right].
\ee
It is now a straightforward exercise to show that the missing enhancement generators can be constructed as,
\be
\left[E_{+-}\right]^I_x= \left[L_L\right]_+\otimes\left[L_R\right]^I_{-x}\qquad\text{and}\qquad \left[E_{-+}\right]^{I}_x= \left[L_L\right]_-\otimes\left[L_R\right]^{I}_{+x}.
\ee

\newpage


%

\end{document}